\documentclass{jpp}

\usepackage{graphicx}
\usepackage{mathrsfs}
\usepackage{amsmath}
\usepackage{natbib}
\usepackage{color}

\ifCUPmtlplainloaded \else
  \checkfont{eurm10}
  \iffontfound
    \IfFileExists{upmath.sty}
      {\typeout{^^JFound AMS Euler Roman fonts on the system,
                   using the 'upmath' package.^^J}%
       \usepackage{upmath}}
      {\typeout{^^JFound AMS Euler Roman fonts on the system, but you
                   dont seem to have the}%
       \typeout{'upmath' package installed. JPP.cls can take advantage
                 of these fonts, if you use 'upmath' package.^^J}%
      }
  \else
  \fi
\fi

\ifCUPmtlplainloaded \else
  \checkfont{msam10}
  \iffontfound
    \IfFileExists{amssymb.sty}
      {\typeout{^^JFound AMS Symbol fonts on the system, using the
                'amssymb' package.^^J}%
       \usepackage{amssymb}%
         \let\leq=\leqslant
         \let\geq=\geqslant
      }{}
  \fi
\fi

\ifCUPmtlplainloaded \else
  \IfFileExists{amsbsy.sty}
    {\typeout{^^JFound the 'amsbsy' package on the system, using it.^^J}%
     \usepackage{amsbsy}}
    {\providecommand\boldsymbol[1]{\mbox{\boldmath $##1$}}}
\fi

\providecommand\bnabla{\boldsymbol{\nabla}}

\newsavebox{\astrutbox}
\sbox{\astrutbox}{\rule[-5pt]{0pt}{20pt}}

\newcommand\Bv{{\bf B}}
\newcommand\Jv{{\bf J}}
\newcommand\uv{{\bf u}}

\newcommand\bv{{\bf b}}

\newcommand\ev{{\bf e}}

\newcommand\bI{{\bf I}}

\newcommand\btau{\boldsymbol{\tau}}
\newcommand\bQ{{\bf Q}}
\newcommand\bQa{{\bf Q}_\alpha}

\newcommand\Omegaca{\Omega_{c\alpha}}

\newcommand\Omegaci{\Omega_{c{\rm i}}}
\newcommand\Omegacp{\Omega_{c{\rm p}}}
\newcommand\ma{m_\alpha}
\newcommand\me{m_{\rm e}}

\newcommand\mpp{m_{\rm p}}

\newcommand\dpp{d_{\rm p}}

\newcommand\nee{n_{\rm e}}

\newcommand\npp{n_{\rm p}}
\newcommand\uav{{\bf u}_\alpha}

\newcommand\ppara{p_\|}
\newcommand\pperp{p_\perp}
\newcommand\pparaa{p_{\|\alpha}}
\newcommand\pperpa{p_{\perp\alpha}}
\newcommand\pparae{p_{\|{\rm e}}}
\newcommand\pperpe{p_{\perp{\rm e}}}

\newcommand\pparap{p_{\|{\rm p}}}
\newcommand\pperpp{p_{\perp{\rm p}}}
\newcommand\Tparap{T_{\|{\rm p}}}

\newcommand\pparaaz{p_{\|\alpha,0}}
\newcommand\pperpaz{p_{\perp\alpha,0}}
\newcommand\pparaez{p_{\|{\rm e},0}}
\newcommand\pperpez{p_{\perp{\rm e},0}}

\newcommand\pparapz{p_{\|{\rm p},0}}
\newcommand\pperppz{p_{\perp{\rm p},0}}
\newcommand\gammapara{\gamma_\|}
\newcommand\gammaperp{\gamma_\perp}

\newcommand\gammaparae{\gamma_{\|{\rm e}}}
\newcommand\gammaperpe{\gamma_{\perp{\rm e}}}

\newcommand\gammaparap{\gamma_{\|{\rm p}}}
\newcommand\gammaperpp{\gamma_{\perp{\rm p}}}
\newcommand\bPi{\boldsymbol{\Pi}}
\newcommand\bPia{\boldsymbol{\Pi}_\alpha}
\newcommand\bPie{\boldsymbol{\Pi}_{\rm e}}

\newcommand\bPip{\boldsymbol{\Pi}_{\rm p}}

\newcommand\bpip{\boldsymbol{\pi}_{\rm p}}

\newcommand\qpara{q_\|}
\newcommand\qperp{q_\perp}
\newcommand\qapara{q_{\|\alpha}}
\newcommand\qaperp{q_{\perp\alpha}}

\newcommand\vth{v_{\rm th}}

\newcommand\betaperpp{\beta_{\perp{\rm p}}}
\newcommand\betaparaz{\beta_{\|,0}}
\newcommand\betaperpz{\beta_{\perp,0}}

\newcommand\betaperpaz{\beta_{\perp\alpha,0}}

\newcommand\betaperpez{\beta_{\perp{\rm e},0}}

\newcommand\betaperppz{\beta_{\perp{\rm p},0}}
\newcommand\tbetaperppz{\widetilde{\beta}_{\perp{\rm p},0}}
\newcommand\FF{{\cal F}}
\newcommand\GG{{\cal G}}
\newcommand\HH{{\cal H}}
\newcommand\tFF{\widetilde{\cal F}}
\newcommand\tGG{\widetilde{\cal G}}
\newcommand\tHH{\widetilde{\cal H}}
\newcommand\tUp{\widetilde{U}'}
\newcommand\omegav{\boldsymbol{\omega}}
\newcommand\eijk{\epsilon_{ijk}}

\newcommand\dQaijkdxk{\frac{\partial\,Q_{\alpha,ijk}^{(n-1)}}{\partial x_k}}
\newcommand\duakdxk{\frac{\partial\,u_{\alpha,k}}{\partial x_k}}
\newcommand\duajdxk{\frac{\partial\,u_{\alpha,j}}{\partial x_k}}
\newcommand\duaidxk{\frac{\partial\,u_{\alpha,i}}{\partial x_k}}
\newcommand\duaydx{\frac{d\,u_{\alpha,y}}{dx}}
\newcommand\duydx{\frac{d\,u_y}{dx}}

\newcommand\Piaij{\Pi_{\alpha,ij}}

\newcommand\Piaxx{\Pi_{\alpha,xx}}
\newcommand\Piayy{\Pi_{\alpha,yy}}
\newcommand\Piazz{\Pi_{\alpha,zz}}

\newcommand\Piij{\Pi_{ij}}
\newcommand\Pixx{\Pi_{xx}}
\newcommand\Piyy{\Pi_{yy}}
\newcommand\Pizz{\Pi_{zz}}
\newcommand\Pixy{\Pi_{xy}}
\newcommand\Pixz{\Pi_{xz}}
\newcommand\Piyz{\Pi_{yz}}

\title[FLR equilibrium configurations with sheared flows]{Finite-Larmor-radius equilibrium and currents of the Earth's flank magnetopause}

\author[S.~S.~Cerri]{S.~S.~Cerri\thanks{Email address for correspondence: scerri@astro.princeton.edu}}
\affiliation{$^1$Department of Astrophysical Sciences, Princeton University, Princeton, NJ 08544, USA}
\pubyear{?}
\volume{?}
\pagerange{?}
\date{?; revised ?; accepted ?. - To be entered by editorial office}
\begin{document}

\maketitle

\begin{abstract}

We consider the one-dimensional equilibrium problem of a shear-flow boundary layer
within an ``extended fluid model'' of plasma that includes the Hall and the electron pressure terms in the Ohm's law, as well as dynamic equations for anisotropic pressure for each species and 
first-order finite Larmor radius (FLR) corrections to the ion dynamics. 
We provide a generalized version of the analytic expressions for the equilibrium configuration given in ~\citet{CerriPOP2013}, highlighting their intrinsic asymmetry due to the relative orientation of the magnetic field $\bv=\Bv/|\Bv|$ and the fluid vorticity $\omegav=\bnabla\times\uv$ (``$\omegav\bv$ asymmetry'').
Finally, we show that FLR effects can modify the Chapman--Ferraro current layer at the flank magnetopause in a way that is consistent with the observed structure reported by \cite{HaalandJGRA2014}.
In particular, we are able to qualitatively reproduce the following key features: (i) the dusk-dawn asymmetry of the current layer, (ii) a double-peak feature in the current profiles, and (iii) adjacent current sheets having thicknesses of several ion Larmor radii and with different current directions.
\end{abstract}

%\begin{PACS}
%\end{PACS}

%\newpage
%\tableofcontents
%\newpage 

\section{Introduction}\label{sec:Intro}

A comprehensive modeling of magnetized plasmas and of their multi-scale 
dynamics is an outstanding challenge in laboratory, astrophysical and space plasma research.
In particular, given that direct numerical simulations are nowadays the main tool to address 
such complex dynamics, finding a compromise between an exhaustive theoretical model 
and its actual implementation represents a major goal for computational plasma physics. 

A kinetic model based on the full Vlasov--Maxwell system of equations 
would need to be solved in a six-dimensional phase space 
(three real-space and three velocity-space dimensions),
resolving length and time scales that typically span over several orders of magnitude. 
For this reason, fully kinetic simulations that adopt realistic parameters 
and/or complex geometries are still far from being realizable because of their colossal computational cost.
Moreover, there is overwhelming difficulty in constructing analytical description of Vlasov equilibria in realistic settings. In fact, the few existing examples typically consider very simplified cases (e.g., uniform and homogeneous magnetic field and/or only periodic functions) and still one cannot fully constrain the resulting velocity profiles beforehand and/or provide those equilibria without appealing to a numerical solution of the problem~\citep[see, e.g.,][]{CaiPOP1990,AtticoPegoraroPOP1999,MahajanHazeltinePOP2000,BobrovaPOP2001,MalaraPRE2018}.

On the other hand, a model based on a fluid treatment such as the 
magnetohydrodynamic (MHD) equations neglects most of the characteristic length and time scales inherent to a kinetic description of the plasma dynamics and only need to be solved in real space.
The MHD description thus represents the simplest viable approach, which nevertheless has led to many fundamental theoretical results~\citep[e.g.,][]{ChapmanFerraro1930,FerraroMNRAS1937,Alfven1942,LustSchluter1954,ChandrasekharPNAS1956,ShafranovJETP1958,GradRMP1960,TaylorPRL1974}.
Furthermore, in the last two decades, we have been able to afford 
well-resolved MHD global simulations providing useful insights~\citep[e.g.,][]{GrothJGR2000,SiscoeGMS2000,JiaJGRA2012,JiaJGRA2015,MerkinJGRA2013,LiuJGRA2015,SorathiaJGRA2017,DongAPJL2017}. 
However, in a real system the nonlinear plasma dynamics would naturally develop small scales and bring the effects associated with the neglected kinetic scales back to light, and so a MHD description eventually breaks down. 
Moreover, accounting for the leading kinetic effects may be necessary already to implement a correct initial plasma equilibrium, in order to avoid uncontrolled and spurious readjustments that can affect the subsequent dynamics or to explain certain features of the system under consideration~\citep[e.g.,][]{HenriPOP2013,CerriPOP2013}.

The fully kinetic and MHD descriptions actually represent the two extremes of a wide variety of plasma models.
There are a large number of approaches that try to bridge the above antipodes in different ways:
from the one side, by simplifying a fully kinetic description based on the dismissal of presumably unimportant effects; from the opposite side, by gradually including more and more kinetic effects within a fluid framework. 
The former class of models are usually referred to as ``reduced-kinetic models'', such as the gyrokinetic (GK)~\citep{BrizardRMP2007} and the hybrid Vlasov-Maxwell (HVM)~\citep{ValentiniJCP2007} approximations; the latter are known as ``extended-fluid models'', in which kinetic effects are gradually included in a fluid description. 
This is the case, for instance, when retaining finite Larmor radius (FLR) corrections~\citep{RobertsPRL1962,MacmahonPOF1965}, or when including the effect of linear Landau damping~\citep{Landau1946} by modeling it with a so-called Landau-fluid (LF) closure~\citep[e.g.,][]{HammettPRL1990}.
These two aspects can also be both included within a single framework, such as in the so-called finite-Larmor-radius Landau-fluid (FLRLF) model~\citep{SulemJPP2015}.
However, within the range of validity defined by each model's assumption (``ordering''), reduced-kinetic models still unavoidably face the curse of high dimensionality, and so extended-fluid models still represent an attractive choice when seeking a compromise between kinetic and fluid descriptions. 

The need to extend a standard fluid description of a collisionless plasma to include at least these effects related to a non-gyrotropic pressure tensor is particularly evident when a sheared flow is present: in the collisionless regime, due to FLR effects, the pressure tensor is indeed strongly coupled to the shear flow and they interact over very short time scales~\citep{CerriMSc2012,CerriPOP2013,CerriPOP2014b,DelSartoPRE2016,DelSartoPPCF2017,DelSartoPegoraroMNRAS2018}.
This is exactly the case of the low-latitude boundary layer (LLBL) between the solar-wind flow and the Earth's magnetosphere, where the velocity shear drives the Kelvin-Helmholtz instability (KHI) that generates the observed large-scale ``MHD'' vortices (see, e.g., \citealt{FaganelloCalifanoJPP2017} and references therein).
In such a region, in addition to the vortex dynamics that naturally develops fluctuations on lengthscales comparable to (or even smaller than) the ion gyroradius $\varrho_i$ (or the ion inertial length $d_i$), the ``large-scale'' equilibrium fields and the sheared flow itself vary over typical lenghtscales $L_0$ that do not exceed the ion characteristic scales by a large amount, and so ``$\varrho_i/L_0$ corrections'' cannot be completely neglected.
So far, such a system has been modeled by means of one-dimensional isotropic MHD equilibrium configurations that ensure the total pressure balance, i.e., a balance between the thermal and magnetic scalar pressures of the two plasmas without involving the properties of the background sheared flow. 
However, as soon as FLR effects and/or the full ion pressure tensor are taken into account, the shear flow properties enter the pressure-balance conditions and the simple isotropic MHD configurations are generally no longer an equilibrium~\citep{CerriMSc2012,CerriPOP2013,CerriPOP2014b}
As a result, the system naturally develops shear-driven anisotropies~\citep[e.g.,][]{DeCamillisPPCF2016,DelSartoPRE2016,DelSartoPegoraroMNRAS2018}.
This is important for (at least) two practical reasons.
First, a difficulty arises when comparing the linear evolution of the KHI using fluid and kinetic models. As discussed in \citet{HenriPOP2013}, in which the same isotropic MHD configuration was adopted as an initial condition for simulations using different plasma models (namely, MHD, two-fluid, PIC-Hybrid and full PIC), it was find that violent and uncontrolled readjustments were either injecting large-amplitude fluctuations in the system~\citep[see also][]{DelSartoPPCF2017} and changing the configuration on top of which the instability develops~\citep[see also][]{NakamuraPOP2010}. 
Therefore, these spurious effects would partially mask the actual kinetic effects on the KHI and make a genuine comparison difficult.  
Secondly, using ten years of observations made by the {\it Cluster} satellites, \citet{HaalandJGRA2014} have recently highlighted that the Earth's magnetopause exhibits a current structure that is more complex than the simple MHD layer described by \citet{ChapmanFerraro1930}, as well as a clear asymmetry between the dusk and the dawn sides. 
In addition to the implications for the current system of a planet magnetosphere, these ion kinetic effects can indeed cause the asymmetric development of KHI at the dawn and the dusk sides of such magnetosphere, as well as other non-ideal effects~\citep[e.g.,][]{NaganoJPP1978,HubaGRL1996,TeradaJGRA2002,NakamuraPOP2010,TaylorAnGeo2012,SundbergJGRA2012,MastersPSS2012,HenriPOP2012,DelamereJGRA2013,ParalNatCo2013,LiljebladJGRA2014,WalshAnGeo2014,HaalandJGRA2014,JohnsonSSRv2014,GingellJGRA2015,GershmanJGRA2015,DeCamillisPPCF2016}.

The aim of the present work is to show how the non-ideal behavior of the Chapman--Ferraro layer could be qualitatively understood in terms of a one-dimensional equilibrium of the shear-flow layer within an extended fluid model that includes first-order ion-FLR corrections.
The great simplicity of the treatment presented here allows to derive analytical equilibrium profiles in which the ion-kinetic effects can be clearly identified. 
Therefore this study is meant to be a first step -- a sort of ``proof of concept'' -- towards the identification of the effects possibly leading to the observed behavior of the low-latitude magnetopause layer, rather than an exhaustive description of the actual system. 
In order to achieve a quantitative modeling of the global magnetopause current system within this (or a more comprehensive) extended-fluid model, a numerical approach to the solution of the full three-dimensional problem would likely be required.

The remainder of this paper is organized as follows. 
In Section~\ref{sect:model} we describe the extended two-fluid (eTF) model of \citet{CerriPOP2013} and we outline the procedure for the derivation of the equilibrium profiles (the actual derivation of a general family of solutions for the shear-flow boundary layer equilibrium is provided in Appendix~\ref{app:generic_equil_derivation}); consequences for shear-flow instabilities, agyrotropy and links to turbulent environments are highlighted in Sections \ref{subsec:timescales} to \ref{subsec:links}. 
In Section~\ref{sec:LLBLcase} we show how these profiles can qualitatively explain the observed non-ideal behavior of the LLBL between the solar wind and the Earth's magnetosphere. 
Finally, in Section~\ref{sec:conclusions} conclusions are drawn.
Additionally, explicit considerations on the symmetries of the FLR expansion and on its convergence to a full pressure tensor case are reported in Appendix~\ref{app:1st-order-FLR} and Appendix~\ref{app:FLR-to-fullPI_convergence}, respectively.

\section{The extended two-fluid (eTF) model}\label{sect:model}

Here, we consider a non-relativistic quasi-neutral proton-electron plasma ($\npp\simeq\nee\equiv n$) in the limit of massless electrons, $\me/\mpp\to0$. The model includes the Hall and the electron pressure terms in the generalized Ohm's law, as well as dynamic equations for the gyrotropic pressures of both species and first-order FLR corrections to the protons' pressure tensor\footnote{We note that in the existing literature the name ``extended MHD'' is sometimes used to describe magneto-hydrodynamic models that include Hall terms and electron inertia effects~\citep[see, e.g.,][]{KimuraMorrisonPOP2014}. Hereafter, we will instead refer to a model as an ``extended fluid model'' when certain kinetic effects, such as, for instance, finite-Larmor-radius contributions and/or linear models of Landau damping, are included within a fluid description.}. 
The fluid hierarchy is closed with a double-adiabatic approximation, i.e., by neglecting the heat fluxes, $\qpara=0$ and $\qperp=0$.
Such assumption is indeed justified within a finite-but-small Larmor radius expansion and on timescales much longer than the ion cyclotron timescale, $\rho/L\sim\omega/\Omega\sim\varepsilon\ll1$, where $\rho$ is the thermal Larmor radius, $L$ is the typical length scale of variation for macroscopic quantities, and $\Omega$ is the cyclotron frequency~\citep[see][for explicit equations and further details about the eTF model ordering]{CerriPOP2013}. In fact, by neglecting gradients in the direction of the magnetic field ($\bv\boldsymbol{\cdot}\bnabla=\nabla_\|=0$; see Appendix~\ref{app:generic_equil_derivation}), the expressions for the perpendicular heat fluxes~\citep[see, e.g.,][]{BraginskiiRPP1965,RamosPOP2008} would give a second-order contribution which is ordered out in the eTF model\footnote{This can be seen also from the point of view of the timescales involved. Let us consider the expressions for the heat fluxes given in \citet{RamosPOP2008}, that in the configuration considered here will reduce to ${\bf q}_\perp = \frac{2p_\perp}{m\Omega}{\bf b}\times\nabla T_\perp$ and ${\bf q}_\| = \frac{p_\perp}{2m\Omega}{\bf b}\times\nabla T_\|$. The timescale on which the divergence of these heat fluxes would contribute on the pressure evolution is thus $\tau_{_{\nabla q}}\sim (L_\perp/\rho)^2\Omega^{-1}\sim\varepsilon^{-2}\Omega^{-1}$ (the timescale for $q_\|$ would actually involve an additional anisotropy correction, $T_\|/T_\perp$, which is not very relevant here). Therefore, the divergence of the heat flux can be neglected with respect to the flow timescale as long as $\varepsilon\sqrt{\beta_\perp}\ll u/v_A$ ($u$ is the typical flow velocity and $v_A$ is the Alfv\'en speed), which is satisfied for the cases under study.}.
In this model, the thermal pressure tensor of the protons and of the electrons, $\bPip$ and $\bPie$ respectively, are written as
%%%%%%%%%%%%%%%%%%%%%%%%%%%%%%%%%%%%%%%%%%%%%%%%%%%%%
 \begin{equation}\label{eq:Pi-proton_def}
  \bPip\,=\,\pparap\bv\bv\,+\,\pperpp\btau\,+\,\bpip^{(1)}\,,
 \end{equation}
 \begin{equation}\label{eq:Pi-electron_def}
  \bPie\,=\,\pparae\bv\bv\,+\pperpe\btau\,,
 \end{equation}
%%%%%%%%%%%%%%%%%%%%%%%%%%%%%%%%%%%%%%%%%%%%%%%%%%%%%
where $\bv\equiv\Bv/|\Bv|$ is the magnetic field unit vector, $\btau\equiv{\bf I}-\bv\bv$ is the projector onto the plane 
perpendicular to $\Bv$, and $\pparaa$ and $\pperpa$ are the gyrotropic thermal pressures of the $\alpha$ species parallel and perpendicular to the magnetic field, respectively~\citep{CGL1956}.
In Eq.~(\ref{eq:Pi-proton_def}), $\bpip^{(1)}$ is a traceless symmetric tensor taking into account first-order FLR corrections to the gyrotropic proton pressure (also known as {\em gyroviscous tensor}).
Neglecting the heat fluxes, a general formulation for the gyroviscous tensor components can be 
written as~\citep{MacmahonPOF1965,SchekochihinMNRAS2010,SulemJPP2015} 
%%%%%%%%%%%%%%%%%%%%%%%%%%%%%%%%%%%%%%%%%%%%%%%%%%%%%
       \begin{equation}\label{eq:pi_ij}
        \pi_{{\rm p},ij}^{(1)}\, =\, 
        	\frac{\pperpp}{4\Omegacp}\Big(\epsilon_{ilm}b_lS_{mk}H_{kj}-H_{ik}\epsilon_{jlm}S_{kl}b_m\Big)\,+\,
	\frac{2(\pperpp-\pparap)}{\Omegacp}\Big(b_iw_j+b_jw_i\Big)\,,
       \end{equation}
%%%%%%%%%%%%%%%%%%%%%%%%%%%%%%%%%%%%%%%%%%%%%%%%%%%%%
where $\Omegacp=eB/\mpp c$ is the proton gyro-frequency, $\epsilon_{ijk}$ is the completely antisymmetric Levi--Civita tensor, and we have introduced 
$S_{ij}\equiv\partial_i u_{{\rm p},j}+\partial_j u_{{\rm p},i}$, $H_{ij}\equiv\delta_{ij}+3b_ib_j$ and
$w_i\equiv\epsilon_{ijk}(\nabla_\|u_{{\rm p},j})b_k$, with $\nabla_\|\equiv\bv\cdot\bnabla$.
Note that the above formulation automatically takes into account for the asymmetry due to the magnetic field direction 
with respect to the vorticity~\citep[see also Appendix~\ref{app:1st-order-FLR} and][for explicit symmetry considerations]{CerriPOP2013}. 

\subsection{Shear-flow layer equilibrium with FLR}

Within this model, we now outline the derivation of equilibrium profiles for a one-dimensional velocity-shear layer separating, for instance, two different plasmas.
The explicit derivation of this class of analytical solutions that generalize the results provided in \citet{CerriPOP2013} and that include a much wider range of configurations of interest for what concerns magnetospheric observations will be provided in Appendix~\ref{app:generic_equil_derivation}. 
The goal is to provide an equilibrium configuration with FLR corrections for the flank magnetopause, and to discuss the implications on the low-latitude boundary layer (LLBL) profiles. 
For the sake of simplicity, here we consider the one-dimensional equilibrium problem, which can be seen as a local approximation of the LLBL.
A global treatment of the magnetospheric structure should take into account curvature terms, as well as possible gradients parallel to the magnetic field and compressible flows. This may need to include additional equilibrium conditions that involve all the gyroviscous components and eventually to go beyond the simple adiabatic FLR treatment presented here by, for instance, including heat fluxes~\citep[see, e.g.,][]{SulemJPP2015,DelSartoPegoraroMNRAS2018}.

We consider a given $x$-dependent incompressible MHD flow in the $y$-$z$ plane, 
%%%%%%%%%%%%%%%%%%%%%%%%%%%%%%%%%%%%%%%%%%%%%%%%%%%%%
\begin{equation}\label{eq:vel_shear_generic}
\uv\,=\,u_y(x)\ev_y\,+\,u_z(x)\ev_z\,,\qquad\bnabla\boldsymbol{\cdot}\uv\,=\,0\,,
\end{equation}
%%%%%%%%%%%%%%%%%%%%%%%%%%%%%%%%%%%%%%%%%%%%%%%%%%%%%
such that it becomes constant at the boundaries (i.e., we consider a localized velocity shear layer). 
The magnetic field also lies on the $y$-$z$ plane, 
%%%%%%%%%%%%%%%%%%%%%%%%%%%%%%%%%%%%%%%%%%%%%%%%%%%%%
\begin{equation}\label{eq:Bfield_generic}
\Bv(x)\,=\,B_y(x)\ev_y\,+\,B_z(x)\ev_z\,.
\end{equation}
%%%%%%%%%%%%%%%%%%%%%%%%%%%%%%%%%%%%%%%%%%%%%%%%%%%%%x 
We further simplify the problem by assuming a polytropic relation for the thermal pressures\footnote{Note that, when heat fluxes are neglected, the natural closure relations for the gyrotropic pressure components would be provided by the double-adiabatic law~\citep{CGL1956} \citep[see, e.g., also][for convenient formulation and extensions]{HauGRL1993,HauPOP2002}. In the case considered here of incompressible flow, no heat fluxes and no gradients parallel to the magnetic field, the double-adiabatic relations and the dynamical pressure equations in the eTF model are equivalent to two different polytropic relations for $p_\|$ and $p_\perp$, namely $\gamma_\perp=2$ and $\gamma_\|=1$~\citep[see, e.g.,][]{CerriMSc2012,CerriPOP2014b,DelSartoPegoraroMNRAS2018}.}. This assumption is not strictly necessary in order to derive the equilibrium, but it is useful for providing density and temperature profiles from the obtained pressure profiles.
In general, the equilibrium for this configuration is found by imposing total pressure balance:
%%%%%%%%%%%%%%%%%%%%%%%%%%%%%%%%%%%%%%%%%%%%%%%%%%%%%
\begin{equation}\label{eq:equil-cond_general}
  \frac{\rm d}{{\rm d}x}\Big[\bPip(x)+\bPie(x)+\bPi_{\rm B}(x)\Big]\,=\,0\,,
\end{equation}
%%%%%%%%%%%%%%%%%%%%%%%%%%%%%%%%%%%%%%%%%%%%%%%%%%%%%
where $\bPi_{\rm B}\equiv (B^2/8\pi){\bf I} -\Bv\Bv$ is the magnetic pressure tensor (${\bf I}$ being the identity tensor).
Within an (anisotropic) MHD model of plasma, the shear-flow does not play a role in the equilibrium profile. 
In fact, when $\bpip^{(1)}$ is neglected, the equilibrium condition for the above configuration simply consists of a balance between the magnetic pressure, $P_B(x)=B^2(x)/8\pi$, and the perpendicular thermal pressures, $P_\perp(x)=\pperpp(x)+\pperpe(x)$. 
In particular, that includes the widely adopted uniform and homogeneous plasma configuration, namely $\pperpa=\pparaaz$, $\pperpa=\pperpaz$, $B_y=B_{0y}$, and $B_z=B_{0z}$, that is not allowed anymore when FLR corrections (or the full pressure-tensor equations) are included in the fluid description~\citep{CerriMSc2012,CerriPOP2013,CerriPOP2014b}.
In general, the solution of the MHD equilibrium condition is completely described by the chosen magnetic profile in (\ref{eq:Bfield_generic}), which determines all the profiles of the other relevant quantities.
Let us now consider the changes of a given MHD equilibrium profile that are induced by a velocity shear of the type described above when first-order FLR corrections are taken into account.
In this case, the only component of $\bpip^{(1)}$ that is relevant to the equilibrium condition is
%%%%%%%%%%%%%%%%%%%%%%%%%%%%%%%%%%%%%%%%%%%%%%%
\begin{equation}\label{eq:pi_xx_equil_generic}
        \pi_{{\rm p},xx}^{(1)}\, =\,  
        -\,\frac{1}{2}\frac{\pperpp}{\Omegacp}
        \left(b_z\frac{{\rm d}\, u_y}{{\rm d}x}\,-\,b_y\frac{{\rm d}\, u_z}{{\rm d}x}\right)\,.
\end{equation}
%%%%%%%%%%%%%%%%%%%%%%%%%%%%%%%%%%%%%%%%%%%%%%%%%%%%%
From (\ref{eq:pi_xx_equil_generic}), one directly identifies the connection between the fluid vorticity,
$\boldsymbol{\omega}\equiv\bnabla\boldsymbol{\times}\uv$, and the magnetic field direction $\bv$, arising as a consequence of the FLR effects:  
%%%%%%%%%%%%%%%%%%%%%%%%%%%%%%%%%%%%%%%%%%%%%%%%%%%%%
\begin{equation}\label{eq:pi_xx_equil-cond_generic_vort}
  \pi_{{\rm p},xx}^{(1)}\, =\,  
    -\,\frac{1}{2}\frac{\pperpp}{\Omegacp}(\bv\boldsymbol{\cdot}\boldsymbol{\omega})\quad\longrightarrow\quad
   \frac{\rm d}{{\rm d}x}\left[
    \left(1-\frac{\mpp c}{eB}\frac{\bv\boldsymbol{\cdot}\boldsymbol{\omega}}{2}\right)\pperpp\,+\,\pperpe\,+\,\frac{B^2}{8\pi}\right]\,=\,0\,,
\end{equation}
%%%%%%%%%%%%%%%%%%%%%%%%%%%%%%%%%%%%%%%%%%%%%%%%%%%%%
where $\omega_y=-u_z'$ and $\omega_z=u_y'$ are the components of the fluid vorticity in our configuration.
Therefore, the FLR corrections give rise to an intrinsic asymmetry in the system's configurations, pressure anisotropy (and most likely also the subsequent dynamics), which depends on the degree of alignment (or anti-alignment) between the flow vorticity and the magnetic field, namely on the sign of $\bv\boldsymbol{\cdot}\boldsymbol{\omega}$.
Such asymmetry has been highlighted in previous numerical simulations and analytical studies~\citep[see, e.g.,][]{NaganoJPP1978,HazeltinePOF1987,CaiPOP1990,HubaGRL1996,RamosPOP2005b,NakamuraPOP2010,HenriPOP2013,CerriPOP2013,DelSartoPRE2016,DelSartoPPCF2017,FranciAIPC2016,ParasharMatthaeusAPJ2016,YangPOP2017,DelSartoPegoraroMNRAS2018}.
We stress, however, that the simple dependence on $\boldsymbol{\omega}$ and $\bv$ in (\ref{eq:pi_xx_equil_generic}) is related to the simplified character of the configuration considered here.

Now assume that $\FF_\perp(x)$, $\GG_\perp(x)$, and $\HH(x)$ are the solutions for the anisotropic MHD equilibrium describing the profiles of the proton perpendicular pressure, $\pperpp=\pperppz\FF_\perp(x)$, of the electron perpendicular pressure, $\pperpe=\pperpez\GG_\perp(x)$, and of the magnetic pressure, $P_B(x)=\frac{B_0^2}{8\pi}\HH(x)$ (here $\pperppz$, $\pperpez$ and $B_0$ are the asymptotic constant values of the pressures and of the magnetic field away from the shear layer, on one of the two sides -- here we do not assume a symmetric shear layer; see Appendix~\ref{app:generic_equil_derivation} for details).
We now seek FLR-corrected equilibrium profiles in the form $\tFF_\perp(x)=\FF_\perp(x)f_\perp(x)$, $\tGG_\perp(x)=\GG_\perp(x)g_\perp(x)$ and $\tHH(x)=\HH(x)h(x)$, where $f_\perp$, $g_\perp$ and $h$ 
are the ``correction functions''. 
By requiring quasi-neutrality and that the MHD profile $\betaperpp(x)$ 
does not change when passing to the corresponding FLR-corrected profile, the solution can be given in term of one function only, i.e.,x $f_\perp(x)=g_\perp(x)=h(x)$ (see Appendix~\ref{app:generic_equil_derivation}):
%%%%%%%%%%%%%%%%%%%%%%%%%%%%%%%%%%%%%%%%%%%%%%%%%%%%%
\begin{equation}\label{eq:f_Blocal_solution}
 f_\perp(x)\,=\,\left\{\frac{\tUp(x)}{2}\,+\sqrt{1+\left(\frac{\tUp(x)}{2}\right)^2\,}\,\right\}^2\,\,.
\end{equation}
%%%%%%%%%%%%%%%%%%%%%%%%%%%%%%%%%%%%%%%%%%%%%%%%%%%%%
where we have defined
%%%%%%%%%%%%%%%%%%%%%%%%%%%%%%%%%%%%%%%%%%%%%%%%%%%%%
\begin{equation}\label{eq:Utilde_def}
 \tUp(x)\,\equiv\,\frac{\tbetaperppz}{2}\,\frac{\mpp\,c}{e\,B_0}\,
 \frac{\FF_\perp(x)}{\HH(x)}\left(\frac{B_{0z}}{B_0}H_z(x)u_y'(x)\,
 -\,\frac{B_{0y}}{B_0}H_y(x)u_z'(x)\right)
\end{equation}
%%%%%%%%%%%%%%%%%%%%%%%%%%%%%%%%%%%%%%%%%%%%%%%%%%%%%
with $\tbetaperppz\equiv\betaperppz/(1+\betaperpz)$ for brevity. 
Note that the solution in (\ref{eq:f_Blocal_solution}) has been obtained taking into account the FLR corrections computed 
with the self-consistent (i.e., FLR-corrected) equilibrium magnetic field profile, $B(x)=B_0\sqrt{\HH(x)f_\perp(x)}$. 
The equilibrium profiles resulting from (\ref{eq:f_Blocal_solution}) are then naturally asymmetric with respect to the sign of $\omegav\boldsymbol{\cdot}\bv$.

\subsection{FLR profiles and approximate kinetic equilibria}

The profiles derived above can be used to initialize the ion distribution function in order to set up an approximate kinetic equilibrium~\citep[see][]{CerriPOP2013}. 
For instance, assuming the inhomogeneity direction to be along $x$, the magnetic field to be in the $z$-direction, $\Bv=B_z(x)\,\ev_z$, and the flow to be along the $y$-axis, $\uv=u_y(x)\,\ev_y$, one obtains the following temperatures:
%%%%%%%%%%%%%%%%%%%%%%%%%%%%%%%%%%%%%%%%%%%%%%%%%%%%%
\begin{equation}\label{eq:Tx_FLRcorrected}
 T_x(x)\,=\,\frac{\pperppz}{n_0}\,\Big(1-\chi(x)\Big)\Big(\FF_\perp(x)\,f_\perp(x)\Big)^{\frac{\gammaperp-1}{\gammaperp}}\,,
\end{equation}
\begin{equation}\label{eq:Tx_FLRcorrected}
 T_y(x)\,=\,\frac{\pperppz}{n_0}\,\Big(1+\chi(x)\Big)\Big(\FF_\perp(x)\,f_\perp(x)\Big)^{\frac{\gammaperp-1}{\gammaperp}}\,,
\end{equation}
\begin{equation}\label{eq:Tz_FLRcorrected}
 T_z\,=\Tparap\,=\,\frac{\pparapz}{n_0}\,\Big(\FF_\perp(x)\,f_\perp(x)\Big)^{\frac{\gammapara-1}{\gammaperp}}\,,
\end{equation}
%%%%%%%%%%%%%%%%%%%%%%%%%%%%%%%%%%%%%%%%%%%%%%%%%%%%%
from which the three thermal velocities, $v_{{\rm th},x}(x)$, $v_{{\rm th},y}(x)$, and $v_{{\rm th},z}(x)$ can be defined. 
The parameter $\chi$ is defined by the first-order FLR correction to the pressure tensor in (\ref{eq:pi_xx_equil-cond_generic_vort}), and provides the agyrotropy of the distribution as a function of the alignment between the flow vorticity, $\omegav$, and the self-consistent FLR-corrected magnetic field. In our transverse case with $\uv=u_y(x)\ev_z$ and $\Bv=B_z(x)\ev_z$, it reads
%%%%%%%%%%%%%%%%%%%%%%%%%%%%%%%%%%%%%%%%%%%%%%%%%%%%%
\begin{equation}\label{eq:FLR-Chi_def}
 \chi(x)\,\equiv\,\frac{1}{2}\frac{\mpp c}{e|\Bv|}\,(\boldsymbol{\omega}\cdot\bv)\,=\,
 \frac{1}{2}\frac{\mpp c}{eB_0}\frac{u_y'(x)}{H_z(x)\sqrt{f_\perp(x)}}\,,
\end{equation}
%%%%%%%%%%%%%%%%%%%%%%%%%%%%%%%%%%%%%%%%%%%%%%%%%%%%%
where $u_y'(x)={\rm d}u_y/{\rm d}x$.
The ``Maxwellian-like'' particle distribution function corresponding to the above profiles reads
%%%%%%%%%%%%%%%%%%%%%%%%%%%%%%%%%%%%%%%%%%%%%%%%%%%%%
\begin{equation}\label{eq:DF-FLRequil}
 F_{\rm M}^{\rm(FLR)}(x,v_x,v_y,v_z)=\frac{(2\pi)^{-3/2}\,n(x)}{\sqrt{T_x(x)\,T_y(x)\,T_z(x)}}\,\exp\left\{-\frac{v_x^2}{2T_x(x)}-\frac{\big(v_y-u_y(x)\big)^2}{2T_y(x)}-\frac{v_z^2}{2T_z(x)}\right\}\,.
\end{equation}
%%%%%%%%%%%%%%%%%%%%%%%%%%%%%%%%%%%%%%%%%%%%%%%%%%%%%

Note that, in the general case, a distribution function reproducing the FLR-corrected profiles would be more complicated, since it may have to give non-diagonal pressure terms. 
Nevertheless, the equilibrium profiles derived from the FLR correction function $f_\perp(x)$ in (\ref{eq:f_Blocal_solution}) still holds for a generic flow and magnetic-field profile (given that they lie in the plane perpendicular to the inhomogeneity direction; see \S~\ref{subsec:assumptions}) and can be used to set up such ``Maxwellian-like'' distributions. 
We stress anyway that a distribution function built in this way is only an approximate kinetic equilibrium, which nevertheless can strongly reduce the spurious fluctuations arising from a readjustment induced by adopting MHD-like equilibrium profiles within a kinetic (or a hybrid-kinetic) framework. 
Unfortunately, exact solutions of the kinetic (or of the hybrid-kinetic) problem usually need to consider simplified configurations, e.g., of the magnetic field, and/or cannot exactly constraint the resulting velocity profiles beforehand~\citep[see, e.g.,][]{CaiPOP1990,AtticoPegoraroPOP1999,MahajanHazeltinePOP2000,BobrovaPOP2001,MalaraPRE2018}. Which solution is better to use clearly depends on the problem under consideration. For instance, in the context of the Earth's flank magnetopause we are dealing with inhomogeneous magnetic field and density profiles (and directions), so the approach presented here is more appropriate for that case.

\subsection{Readjustment timescale of unbalanced equilibria}\label{subsec:timescales}

As mentioned in the Introduction, taking into account the leading kinetic effects (such as the above first-order ion-FLR correction) may be necessary already at the level of the initial plasma configuration. 
In fact, adopting an ideal MHD initial equilibrium in a kinetic framework will result in a quick readjustment and in the development of spurious large-amplitude fluctuations~\citep[see][]{HenriPOP2013,CerriPOP2013}. 

%================================
\begin{figure}
\centering
 \includegraphics[width=0.49\textwidth]{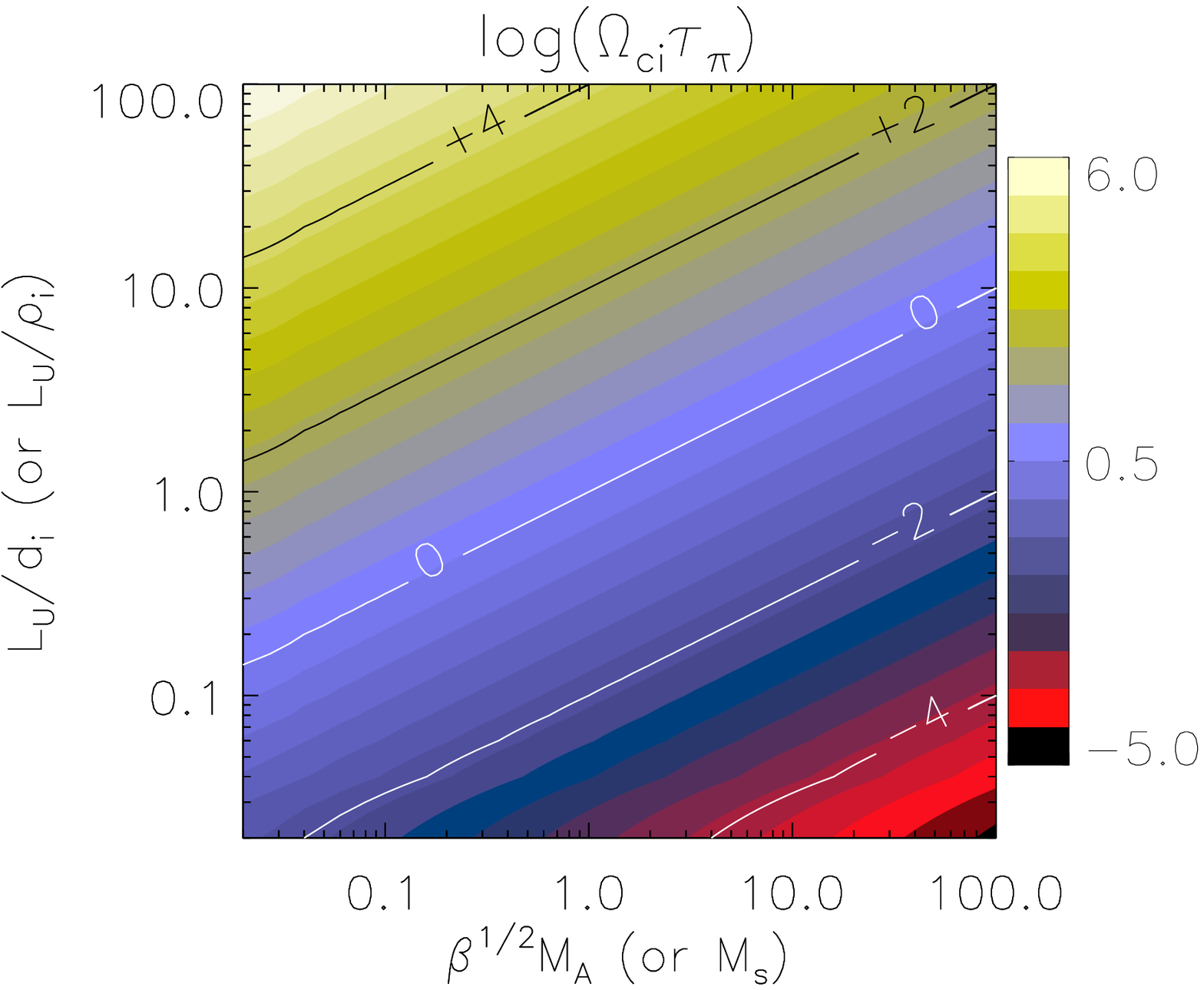}
 \includegraphics[width=0.49\textwidth]{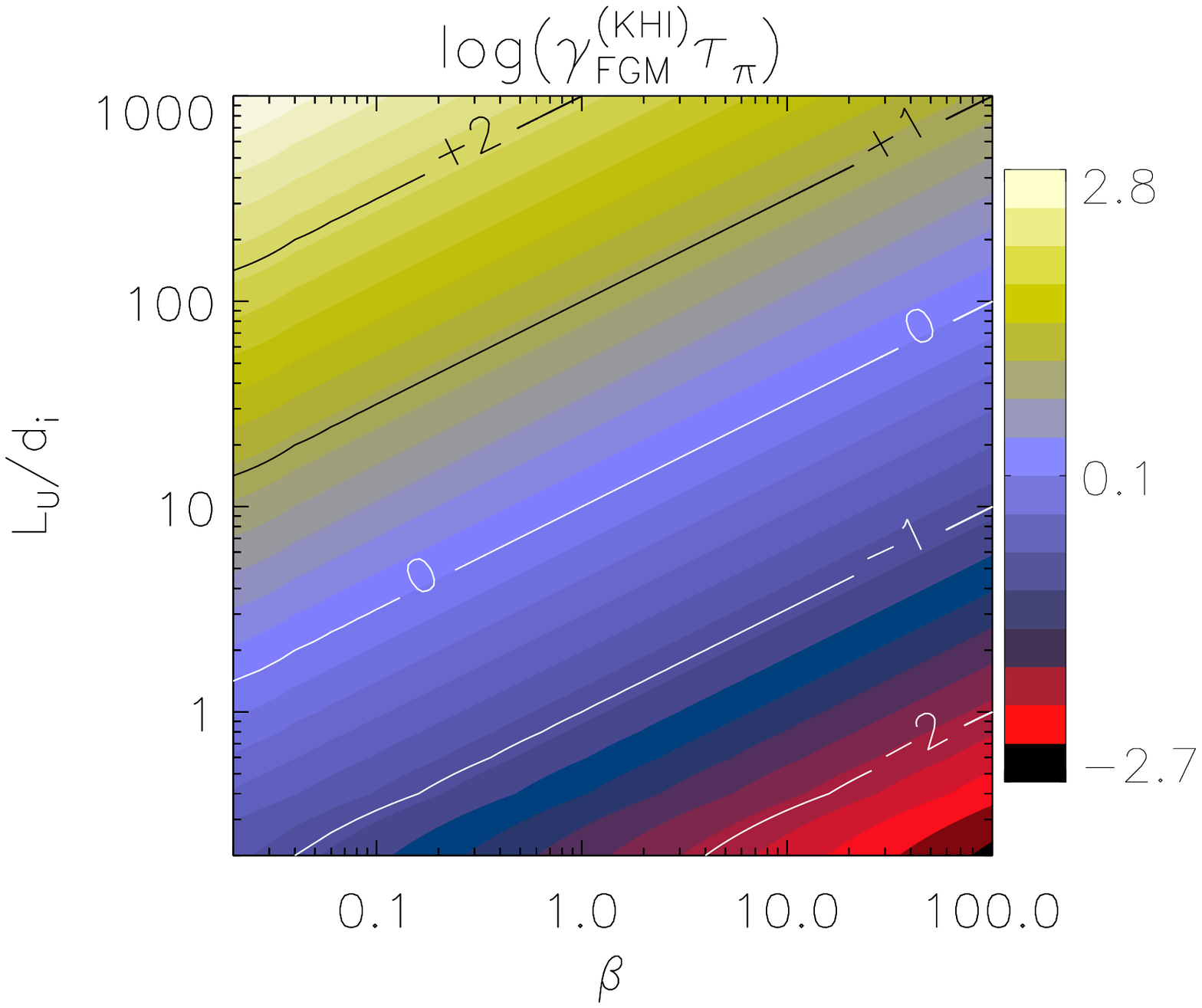}
 \caption{Left panel: iso-surfaces of $\log(\Omegaci\tau_\pi)$ in the $L_u/d_i$ versus $\beta^{1/2}M_A$ plane. The same iso-surfaces apply to the $L_u/\rho_i$ versus $M_s$ plane ($M_s\equiv u_0/c_s$ is the Mach number, $c_s$ being the sound speed). Right panel: iso-surfaces of $\log(\gamma_{_{\rm FGM}}^{\rm(KHI)}\tau_\pi)$ in the $L_u/d_i$ versus $\beta$ plane.}
 \label{fig:FLR_FLR-KHI_timescales}
\end{figure} 
%================================

When MHD equilibria are employed in kinetic simulations where a sheared flow is present, the unbalanced leading ion-FLR corrections will induce a readjustment on timescales $\tau_\pi$ of the order\footnote{Here we are assuming that the corresponding electron-FLR corrections are negligible compared to those of the ions. This assumption may break down for $\beta_{e,\perp}\sim\frac{m_i}{m_e}\beta_{i,\perp}\gg\beta_{i,\perp}$.}
%%%%%%%%%%%%%%%%%%%%%%%%%%%%%%%%%%%%%%%%%%%%%%%%%%%%%
\begin{equation}\label{eq:tau_pi}
 \tau_\pi^{-1}\,\sim\,\beta_{i,\perp}^{-1/2}\,M_A\left(\frac{\rho_i}{L_u}\right)^2\Omega_{c,i}\,\sim\,\beta_{i,\perp}^{1/2}\,M_A\left(\frac{d_i}{L_u}\right)^2\Omega_{c,i}\,,
\end{equation}
%%%%%%%%%%%%%%%%%%%%%%%%%%%%%%%%%%%%%%%%%%%%%%%%%%%%%
where $M_A\equiv u_0/v_A$ and $L_u$ are the Alfv\'enic Mach number and lengthscale of the background shear flow. 
It may be useful to compare this readjustment timescale with the growth rate of the fastest-growing-mode (FGM) for the Kelvin-Helmholtz instability,
%%%%%%%%%%%%%%%%%%%%%%%%%%%%%%%%%%%%%%%%%%%%%%%%%%%%%
\begin{equation}\label{eq:gamma_FGM_KHI}
 \gamma_{_{\rm FGM}}^{\rm(KHI)}\,\sim\,\frac{1}{4}k_{_{\rm FGM}}u_0\,
 \sim\,0.1\,\beta_{i,\perp}^{-1/2}\,M_A\left(\frac{\rho_i}{L_u}\right)\Omega_{c,i}\,
 \sim\,0.1\,M_A\left(\frac{d_i}{L_u}\right)\Omega_{c,i}\,,
\end{equation}
%%%%%%%%%%%%%%%%%%%%%%%%%%%%%%%%%%%%%%%%%%%%%%%%%%%%%
where we have used the relation $k_{_{\rm FGM}}L_u\sim0.4$ derived in the compressible MHD limit~\citep[see][and references therein]{FaganelloCalifanoJPP2017}. Therefore, the effects of such readjustment on the KHI growth are of order
%%%%%%%%%%%%%%%%%%%%%%%%%%%%%%%%%%%%%%%%%%%%%%%%%%%%%
\begin{equation}\label{eq:gamma_KHI-tau_pi}
 \gamma_{_{\rm FGM}}^{\rm(KHI)}\tau_\pi\,\sim\,0.1\left(\frac{L_u}{\rho_i}\right)
 \sim\,0.1\,\beta_{i,\perp}^{-1/2}\left(\frac{L_u}{d_i}\right)\,,
\end{equation}
%%%%%%%%%%%%%%%%%%%%%%%%%%%%%%%%%%%%%%%%%%%%%%%%%%%%%
which is typically smaller than (or order of) unity for the magnetopause case, meaning that any readjustment happens faster than the instability itself and therefore will strongly change the equilibrium on top of which the KHI develops. 

A sketch of the behavior of timescales in (\ref{eq:tau_pi}) and (\ref{eq:gamma_KHI-tau_pi}) with respect to the relevant parameters is provided in Fig.~\ref{fig:FLR_FLR-KHI_timescales}. MHD-like behavior is recovered in the parameter space denoted by yellow/white colors.

\subsection{Sustainability of pressure agyrotropy}\label{subsec:agyrotropy}

An interesting feature of the interaction between the pressure tensor and a sheared flow is the sustainability and/or the generation of pressure ``agyrotropy''~\citep{CerriPOP2013,CerriPOP2014b,DelSartoPRE2016,DelSartoPPCF2017}. 
This means that, in addition to the typical pressure anisotropy with respect to the magnetic-field direction that is typical of collisionless plasmas ($p_\perp\neq p_\|$), now additional pressure anisotropy can be present in the plane perpendicular to $\Bv$, e.g., $p_{\perp,1}\neq p_{\perp,2}\neq p_\|$, where ($\ev_{\perp,1}$, $\ev_{\perp,2}$, $\ev_\|$) is any orthogonal basis within which the pressure tensor is diagonal and where $\ev_{\perp,1}$ and $\ev_{\perp,2}$ define the plane perpendicular to the magnetic field. 
In this section we analyze this aspect in terms of equilibrium configurations and their corresponding agyrotropy. 
However, we stress that this feature has consequences in the dynamics of a collisionless plasma as well, e.g., modifying linear properties of perturbations~\citep[e.g.,][]{DelSartoPRE2016,DelSartoPPCF2017}, enhancing the kinetic activity related to vorticity, current sheets, reconnection and energy transfer in turbulence~\citep[e.g.,][]{GrecoPRE2012,ServidioPRL2012,ServidioAPJL2014,YangPOP2017}, and possibly affecting the regulation of anisotropies in accretion disks~\citep[e.g.,][]{KunzPRL2016}.

%================================
\begin{figure}
\centering
 \includegraphics[width=0.75\textwidth]{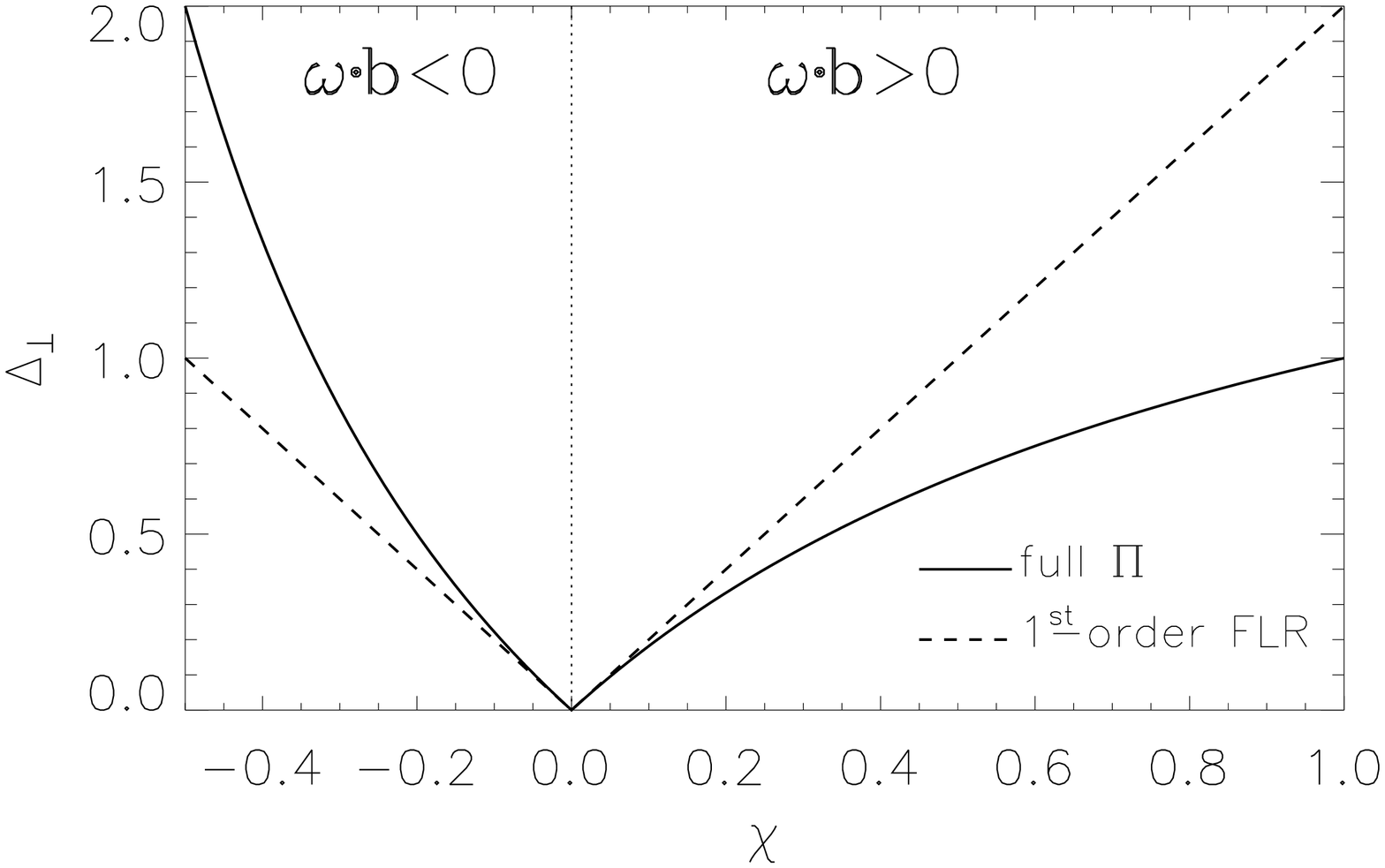}
 \caption{Pressure anisotropy in the plane perpendicular to the magnetic field direction, $\Delta_\perp$, versus $\chi\equiv\boldsymbol{\omega}\cdot\bv/2\Omegaci$ obtained from $1^{\rm st}$--order FLR corrections (dashed line, equation \ref{eq:agyrotropy}) and from the full-pressure tensor equation (continuous line, equation \ref{eq:agyrotropy_fullPI}). Positivity of pressure from the full-$\Pi$ treatment requires $\chi\geq-1/2$~\citep[see][]{CerriPOP2014b}, while the FLR treatment holds for $|\chi|\ll1$.}
 \label{fig:FLR_fullPI_anisotropy}
\end{figure} 
%================================

In our configuration it is easy to show that the FLR effects introduce an {\em agyrotropy}, $\Delta_\perp$, i.e., an anisotropy in the plane perpendicular to the magnetic field~\citep[see, e.g.,][for a general formulation]{ScudderDaughtonJGRA2008}, given by
%%%%%%%%%%%%%%%%%%%%%%%%%%%%%%%%%%%%%%%%%%%%%%%%%%%%%
\begin{equation}\label{eq:agyrotropy}
 \Delta_\perp\,=\,\frac{|\boldsymbol{\omega}\cdot\bv|}{\Omega_{ci}}\,
 \equiv|2\,\chi|\,.
\end{equation}
%%%%%%%%%%%%%%%%%%%%%%%%%%%%%%%%%%%%%%%%%%%%%%%%%%%%%
Since only the first-order FLR corrections have been retained in the present description, only small deviations from gyrotropy are correctly described in this case, i.e., the condition $|\chi|=|\boldsymbol{\omega}\cdot\bv/2\Omegaci|\ll1$ should hold. 
Also, in this approximation the equilibrium exhibits an asymmetry with respect to the sign of $\boldsymbol{\omega}\cdot\bv$, but $\Delta_\perp$ does not. 
In order to have such asymmetry in the agyrotropy, next-order corrections or the full pressure tensor must be retained. In the latter case, the agyrotropy in the plane perpendicular to $\Bv$ would be~\citep{CerriPOP2014b} 
%%%%%%%%%%%%%%%%%%%%%%%%%%%%%%%%%%%%%%%%%%%%%%%%%%%%%
\begin{equation}\label{eq:agyrotropy_fullPI}
 \Delta_\perp\, =\,\left|\frac{2\,\chi}{1+\chi}\right|\,,
\end{equation}
%%%%%%%%%%%%%%%%%%%%%%%%%%%%%%%%%%%%%%%%%%%%%%%%%%%%%
where the condition $\chi\geq-1/2$ must hold because of the positivity constraint on pressure.

In Fig.~\ref{fig:FLR_fullPI_anisotropy} we report a comparison between the pressure anisotropy in the plane perpendicular to the magnetic field, $\Delta_\perp$, as a function of the parameter $\chi$, obtained via the full pressure-tensor equation~\citep{CerriPOP2014b} and via first-order FLR corrections. 

\subsection{A broader view: relevance to other instabilities and turbulent environments}\label{subsec:links}

As we will show in \S~\ref{sec:LLBLcase}, the main consequences related to the ion-FLR effects reported in this paper have a direct effect in the current system of a planetary magnetopause. 
Moreover, these ion-kinetic effects can cause the asymmetric development of KHI at the dawn and the dusk sides of such magnetosphere, as well as other non-ideal features~\citep[see, e.g.,][]{NaganoJPP1978,HubaGRL1996,TeradaJGRA2002,NakamuraPOP2010,TaylorAnGeo2012,SundbergJGRA2012,MastersPSS2012,HenriPOP2012,DelamereJGRA2013,ParalNatCo2013,LiljebladJGRA2014,WalshAnGeo2014,HaalandJGRA2014,JohnsonSSRv2014,GingellJGRA2015,GershmanJGRA2015,DeCamillisPPCF2016}.
However, ion-FLR effects and their relations with anisotropy, vorticity and current sheets can have implications on a wide variety of astrophysical and space scenarios.

In fact there are further shear-driven instabilities that may also get relevant feedback from anisotropy (and agyrotropy) developed (or sustained) by the underlying shear flow within a kinetic description such as, for instance, for the case of magneto-rotational instability (MRI) in accretion disks~\citep[e.g.,][]{FerraroAPJ2007,RiquelmeAPJ2012,KunzPRL2016,SquireJPP2017}. 
Furthermore, ion-kinetic effects such as FLR and pressure-tensor dynamics can affect anisotropy-driven instabilities themselves~\citep[e.g.,][]{SchekochihinMNRAS2010,RosinMNRAS2011,SarratEPL2016,SquirePRL2017}, which are relevant, e.g., in the evolution of the solar wind~\citep[e.g.,][]{HellingerGRL2006,TeneraniAPJ2017,YoonRMPP2017} and in magnetic reconnection~\citep[e.g.][]{SchoefflerAPJ2011,CassakPOP2015}. 

Finally, current sheets and the associated reconnection processes are fundamental ingredients of turbulent plasmas~\citep[e.g.][]{MatthaeusLamkinPOF1986,BiskampBOOK2008,ServidioPOP2010,ServidioNPG2011,LazarianPOP2012,KarimabadiSSR2013,ServidioJPP2015,FranciAIPC2016,CerriJPP2017}.
In this context, currents and coherent structures are typically related to simultaneous enhancement of vorticity, kinetic activity, turbulent transfer and dissipation~\citep[e.g.,][]{ServidioPRL2012,ServidioAPJL2014,KarimabadiPOP2013,ValentiniPOP2014,ValentiniNJP2016,WanPRL2015,FranciAIPC2016,ParasharMatthaeusAPJ2016,YangPOP2017,GroseljAPJ2017,CamporealePRL2018,Sorriso-Valvo-etal2018}. 
Furthermore, reconnection/structures have been recently proved to enhance/trigger the kinetic turbulent cascades in real space~\citep{CerriCalifanoNJP2017,FranciAPJL2017,CamporealePRL2018} and also to be related to simultaneous velocity space cascades~\citep{ServidioPRL2017,CerriAPJL2018,Pezzi2018}.
These reconnecting current sheets and the resulting magnetic structures are quasi-equilibrium pressure-balanced structures with embedded sheared flows even within a turbulent environment~\citep[see e.g.,][]{CerriCalifanoNJP2017}. 
Therefore ion-kinetic effects such as FLR contributions (or the full pressure-tensor; see \citet{CerriPOP2014b}, \citet{YangPOP2017} and \citet{DelSartoPegoraroMNRAS2018}) may play a relevant role in the complex interplay between currents, vorticity, reconnection, non-Maxwellian features, velocity-space cascades and dissipation in turbulent plasmas.

\section{Application to the LLBL of the Earth's magnetopause}\label{sec:LLBLcase}

Let us now consider an explicit application to the LLBL of the Earth's magnetopause, the goal being to show that the observed deviations from the ideal Chapman--Ferraro current system highlighted in \citet{HaalandJGRA2014} can be qualitatively explained with the ions FLR corrections. 
We want to stress that this is not meant to be a quantitative explanation of the observed profiles, since also 3D geometry and other effects may contribute to the actual profiles. 
In what follows, equations are normalized to the proton mass, inertial length and cyclotron frequency ($\mpp$, $\dpp$ and $\Omegacp$, respectively), and the Alfv\'en speed ($v_A$).

We consider a local one-dimensional model the LLBL region in which the inhomogeneity direction ($x$) is perpendicular to the plane ($yz$) where both the flow and the magnetic field lie. 
Typically, hyperbolic tangent give a reasonably realistic modeling of the flow,
%%%%%%%%%%%%%%%%%%%%%%%%%%%%%%%%%%%%%%%%%%%%%%%%%%%%%
\begin{equation}\label{eq:Uy_equil}
 u_y(x)\,=\,u_0\sin\phi\,\tanh\left(\frac{x-x_{u,0}}{L_u}\right)\,,
\end{equation}
%%%%%%%%%%%%%%%%%%%%%%%%%%%%%%%%%%%%%%%%%%%%%%%%%%%%%
%%%%%%%%%%%%%%%%%%%%%%%%%%%%%%%%%%%%%%%%%%%%%%%%%%%%%
\begin{equation}\label{eq:Uz_equil}
  u_z(x)\,=\,u_0\cos\phi\,\tanh\left(\frac{x-x_{u,0}}{L_u}\right)\,,
\end{equation}
%%%%%%%%%%%%%%%%%%%%%%%%%%%%%%%%%%%%%%%%%%%%%%%%%%%%%
where $\phi$ is the angle between the $z$-axis and the plane where the sheared flow velocity lies, and of the magnetic field,
%%%%%%%%%%%%%%%%%%%%%%%%%%%%%%%%%%%%%%%%%%%%%%%%%%%%%
\begin{equation*}
  B_y(x)\,=\, B_0\Bigg\{\frac{B_G}{B_0}\sin\vartheta\,\left[1\,+\,\frac{\Delta B_\|}{2\,B_G}\left(1-\tanh\Big(\frac{x-x_{B,0}}{L_B}\Big)\right)\right]\,+
\end{equation*}
\begin{equation}\label{eq:By_MHDequil}
 \qquad\quad\,+\,\,\frac{\Delta B_\perp}{2B_0}\cos\vartheta\left[1+\tanh\Big(\frac{x-x_{B,0}}{L_B}\Big)\right]\Bigg\}\,,
\end{equation}
%%%%%%%%%%%%%%%%%%%%%%%%%%%%%%%%%%%%%%%%%%%%%%%%%%%%%
%%%%%%%%%%%%%%%%%%%%%%%%%%%%%%%%%%%%%%%%%%%%%%%%%%%%%
\begin{equation*}
  B_z(x)\,=\, B_0\Bigg\{\frac{B_G}{B_0}\cos\vartheta\,\left[1\,+\,\frac{\Delta B_\|}{2\,B_G}\left(1-\tanh\Big(\frac{x-x_{B,0}}{L_B}\Big)\right)\right]\,-
\end{equation*}
\begin{equation}\label{eq:Bz_MHDequil}
 \qquad\quad\,-\,\,\frac{\Delta B_\perp}{2B_0}\sin\vartheta\left[1+\tanh\Big(\frac{x-x_{B,0}}{L_B}\Big)\right]\Bigg\}\,,
\end{equation}
%%%%%%%%%%%%%%%%%%%%%%%%%%%%%%%%%%%%%%%%%%%%%%%%%%%%%
where $B_0=\sqrt{B_G^2+\Delta B_\perp^2}$ and $\vartheta$ is the angle between the $z$-axis and the magnetic field at $x\to-\infty$\footnote{The corresponding angle $\varphi$ between the $z$-axis and $\Bv$ at $x\to\infty$ is related to $\vartheta$ and $\Delta B_\perp$ by $\tan\varphi=(\tan\vartheta+\Delta B_\perp/B_G)/(1-\tan\varphi\,\Delta B_\perp/B_G)$, and $\varphi=\vartheta$ when $\Delta B_\perp=0$.}. 
The above magnetic profile accounts both for variations that are purely in magnitude, through $\Delta B_\|$, and for rotations (magnetic shear) of the magnetic-field direction, through $\Delta B_\perp$~\citep[see, e.g.,][for the effects of $\Delta B_\perp$ on KHI at the Earth's magnetospheric flanks]{Fadanelli2018}.
Note that usually $x_{u,0}=x_{B,0}$ and $L_u=L_B$ are assumed in numerical simulations~\citep[see, e.g.,][]{MiuraJGR1987,FujimotoTerasawaJGR1995,OttoJGR2000,NykyriOttoAnGeo2004,NakamuraFujimotoGRL2005,FaganelloPRL2008,FaganelloEPL2012,PalermoJGRA2011,TeneraniPPCF2011}. 
However, recent satellite measurements have shown that the magnetic (and density) profiles can be slightly shifted 
with respect to the velocity shear and/or that the shear length-scales of these quantities may differ, 
i.e., $x_{u,0}\neq x_{n,0}$ and/or $L_u\neq L_n$~\citep{FoullonJGRA2008,HaalandJGRA2014,RossiPhDthesis2015}.
This idea has been also recently implemented in numerical simulations in order to explain some observational features~\citep{RossiPhDthesis2015,LeoryKeppensPOP2017}. 
Therefore, here we also take into account these features.
For a magnetic profile as in (\ref{eq:By_MHDequil})-(\ref{eq:Bz_MHDequil}) the MHD magnetic pressure function, $\HH$, is given by
%%%%%%%%%%%%%%%%%%%%%%%%%%%%%%%%%%%%%%%%%%%%%%%%%%%%%
\begin{equation}\label{eq:HH_MHDequil}
 \HH(x)\, =\,\frac{B_G^2}{B_0^2}\Bigg\{ \left[1+\frac{\Delta B_\|}{2\,B_G}\left(1-\tanh\Big(\frac{x-x_{B,0}}{L_B}\Big)\right)\right]^2
 +\,\frac{\Delta B_\perp^2}{4B_G^2}\left[1+\tanh\Big(\frac{x-x_{B,0}}{L_B}\Big)\right]^2\Bigg\}\,,
\end{equation}
%%%%%%%%%%%%%%%%%%%%%%%%%%%%%%%%%%%%%%%%%%%%%%%%%%%%%
and the corresponding MHD thermal profiles are obtained in terms of
%%%%%%%%%%%%%%%%%%%%%%%%%%%%%%%%%%%%%%%%%%%%%%%%%%%%%
\begin{equation*}
 \FF_\perp(x)\,=\,\GG_\perp(x)\,=\,
 1\,+\,\frac{\Delta B_\perp^2}{\betaperpz\,B_0^2}\,
 -\,\frac{B_G\,\Delta B_\|}{\betaperpz\,B_0^2}\left[1-\tanh\Big(\frac{x-x_{B,0}}{L_B}\Big)\right]\,-
\end{equation*}
\begin{equation}\label{eq:FF_MHDequil}
 -\,\frac{\Delta B_\|^2}{4\,\betaperpz\,B_0^2}\left[1-\tanh\Big(\frac{x-x_{B,0}}{L_B}\Big)\right]^2\,
 -\,\frac{\Delta B_\perp^2}{4\,\betaperpz\,B_0^2}\left[1+\tanh\Big(\frac{x-x_{B,0}}{L_B}\Big)\right]^2\,,
\end{equation}
%%%%%%%%%%%%%%%%%%%%%%%%%%%%%%%%%%%%%%%%%%%%%%%%%%%%%
where $\betaperpz\,B_0^2=2\,P_{\perp,0}\equiv2\,(\pperppz+\pperpez)$ and the positivity condition on pressure (see (\ref{app:eq:HH-FF_implicit}) in \S~\ref{app:subsec:FLRequil_generic}) here reads as
%%%%%%%%%%%%%%%%%%%%%%%%%%%%%%%%%%%%%%%%%%%%%%%%%%%%%
\begin{equation}\label{eq:FF_MHDequil_positivity}
B_G\,\Delta B_\|\,+\,\frac{\Delta B_\|^2}{2}\, \leq \, P_{\perp,0}\,+\,\frac{\Delta B_\perp^2}{2}\,.
\end{equation}
%%%%%%%%%%%%%%%%%%%%%%%%%%%%%%%%%%%%%%%%%%%%%%%%%%%%%
The FLR corrections to the above MHD profiles are then given in terms of
%%%%%%%%%%%%%%%%%%%%%%%%%%%%%%%%%%%%%%%%%%%%%%%%%%%%%
\begin{equation*}
 \tUp(x)\,=\, \frac{\tbetaperppz}{2}\,\frac{u_0}{B_0\,L_u}\,\,\frac{\FF_\perp(x)}{\HH(x)}\,\cosh^{-2}\left(\frac{x-x_{u,0}}{L_u}\right)\,\times\,,
\end{equation*}
\begin{equation*}
\times\Bigg\{\frac{B_G}{B_0}\left[1+\frac{\Delta B_\|}{2\,B_G}\left(1-\tanh\Big(\frac{x-x_{B,0}}{L_B}\Big)\right)\right]\sin(\phi-\vartheta)\,-
\end{equation*}
\begin{equation}\label{eq:Utildeprimo_KHequil}
-\,\frac{\Delta B_\perp}{2B_0}\left[1+\tanh\Big(\frac{x-x_{B,0}}{L_B}\Big)\right]\cos(\phi-\vartheta)\Bigg\}\,\,,
\end{equation}
%%%%%%%%%%%%%%%%%%%%%%%%%%%%%%%%%%%%%%%%%%%%%%%%%%%%%
which is again related to the sign of the scalar product between the fluid vorticity and the magnetic field through the $\sin(\phi-\vartheta)$ and $\cos(\phi-\vartheta)$ coefficients.

\subsection{Current profiles at the Earth's flank magnetopause: an example}
%--------------------------------
\begin{table}
 \center
 \begin{tabular}{cccc|ccccccc|ccccccccc|cccccccc}
 \hline
 \hline
 \multicolumn{3}{c}{\,} & & & \multicolumn{5}{c}{flow parameters} & & & \multicolumn{7}{c}{magnetic field parameters} & & & \multicolumn{7}{c}{plasma thermal parameters}\\
 %\hline
\multicolumn{3}{c}{\,} & & & & & & & & & & & & & & & & & & & & & & & &\\
    \multicolumn{3}{c}{case} & & & $u_0$ & & $L_u$  & & $x_{u,0}$ & & & $\Delta B_\|$ & & $\Delta B_\perp$ & & $L_B$ & & $x_{B,0}$ & & & $\betaperpz$ & & $\betaparaz$ & & $\gammaperp$ & & $\gammapara$ \\
   \hline
  & & & & & & & & & & & & & & & & & & & & & & & & & &\\
  & A & & & & $\pm1$ & & $2$ & & $\pm1$ & & & $0.5$ & & $0.6$ & & $6$ & & $0$ & & & $2$ & & $2$ & & $2$ & & $1$\\
  & & & & & & & & & & & & & & & & & & & & & & & & & &\\
  & B & & & & $\pm2$ & & $2$ & & $\pm1$ & & & $0.7$ & & $0.7$ & & $6$ & & $0$ & & & $4$ & & $4$ & & $2$ & & $1$\\
  & & & & & & & & & & & & & & & & & & & & & & & & & &\\
  \hline
  \hline
 \end{tabular}
   \caption{Summary of the parameters used for profiles in Fig.~\ref{fig:Jprofiles_FLR}. All the parameters are normalized with respect to quantities characteristic of the SW region: flow speed is in $v_{\rm A}^{\rm(SW)}$ units, lengths are in $d_{\rm i}^{\rm(SW)}$ units and magnetic field variations are in $B_0^{\rm(SW)}=1$ units (from which $B_G=\sqrt{1-\Delta B_\perp^2}$ follows). The plus and minus sign in $u_0$ and in $x_{u,0}$ are for the dusk and for the dawn side, respectively. We also remind the reader that $\vartheta=0$ and $\phi=\pi/2$ in both cases.}
   \label{tab:tab1}
\end{table}
%--------------------------------
%================================
\begin{figure}
\centering
 \includegraphics[width=0.496\textwidth]{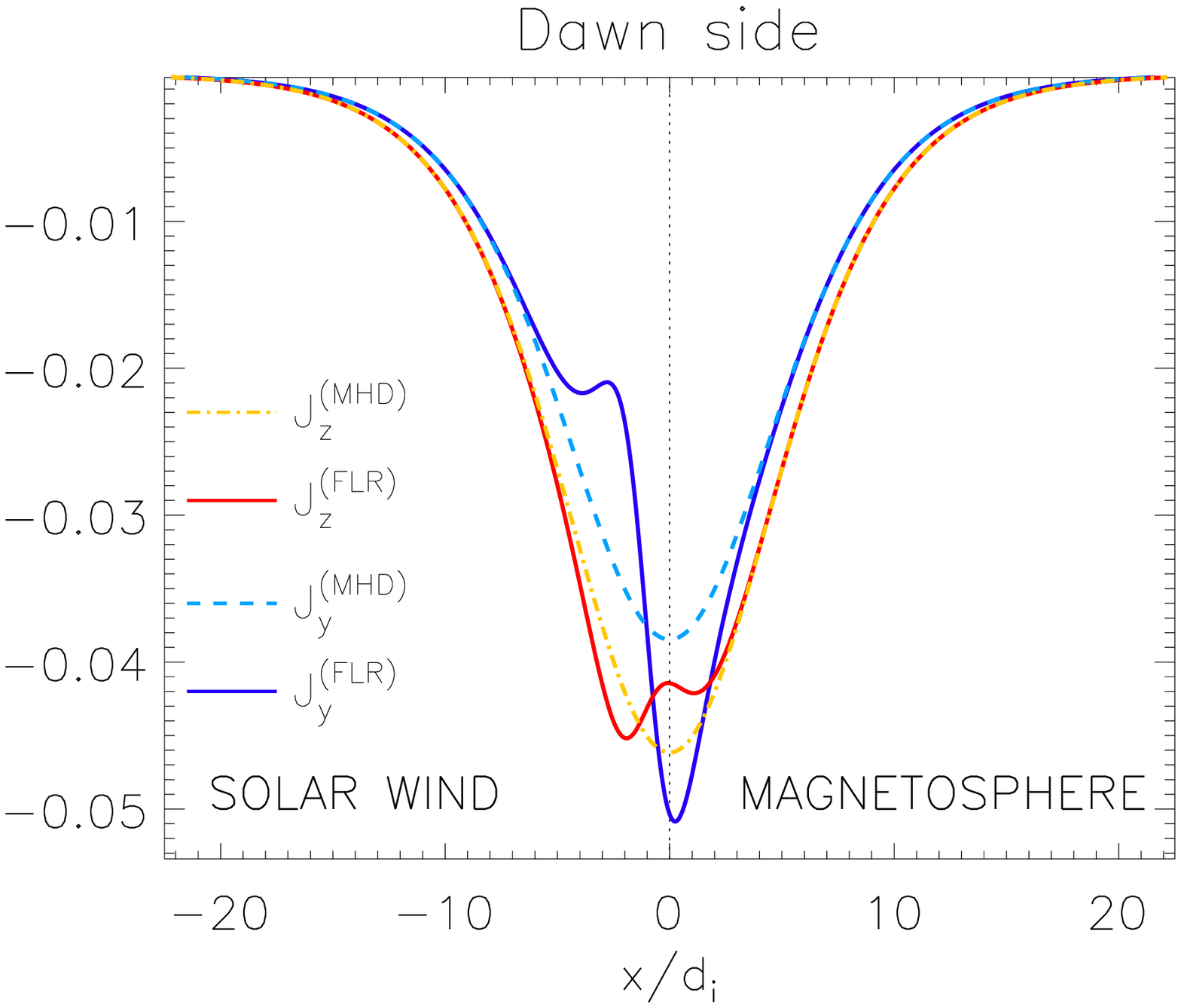}
 \includegraphics[width=0.496\textwidth]{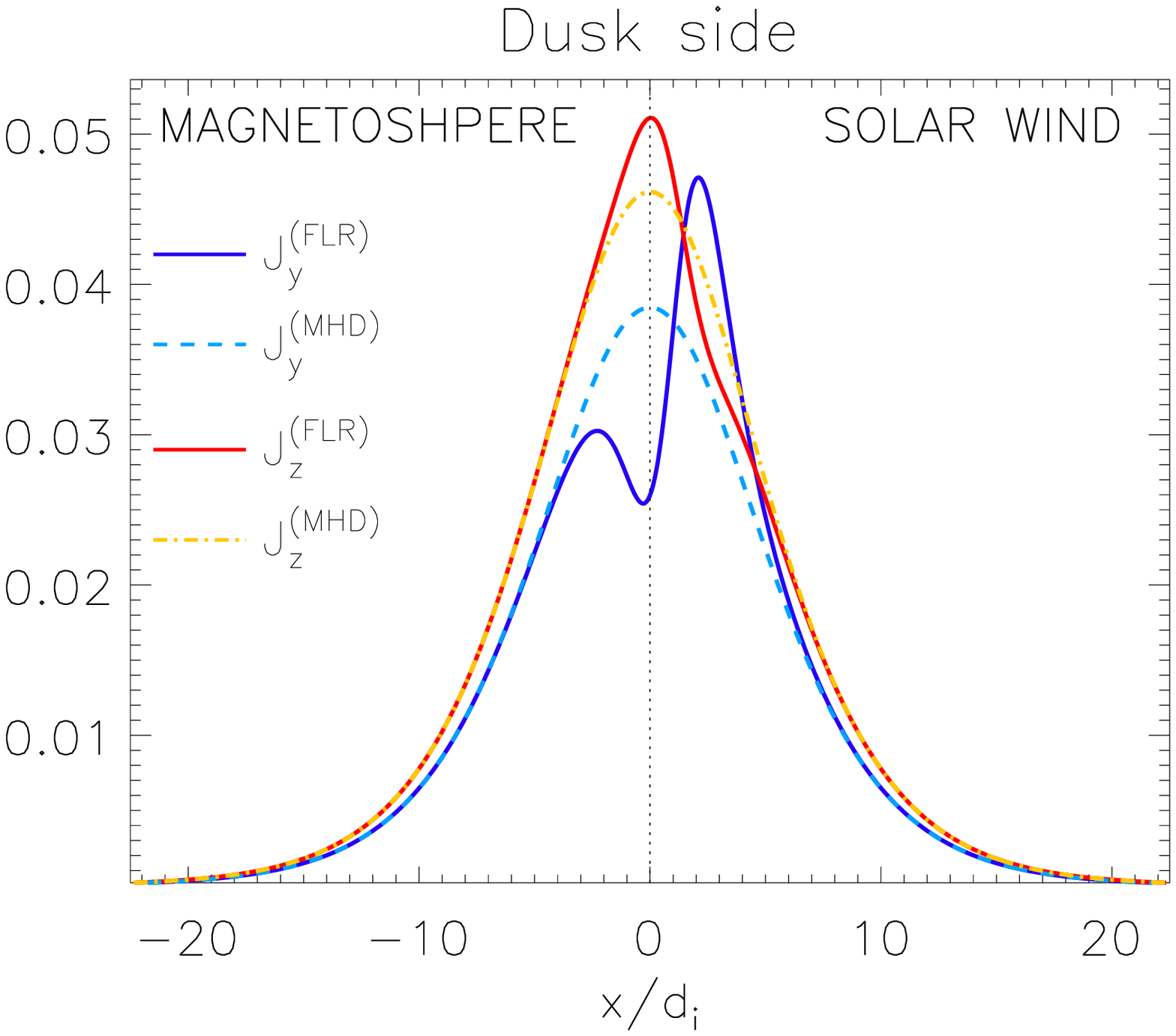}
 \includegraphics[width=0.496\textwidth]{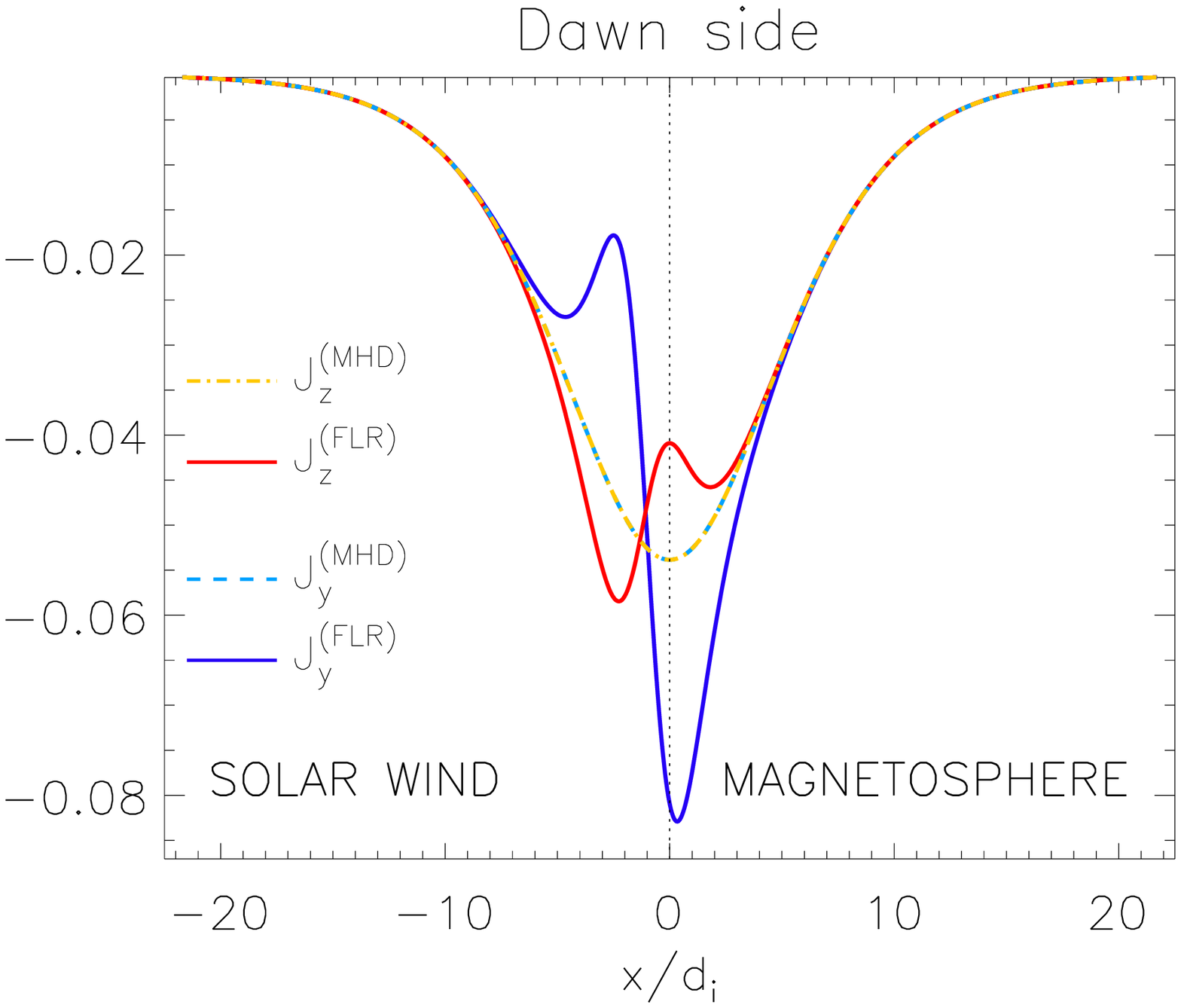}
 \includegraphics[width=0.496\textwidth]{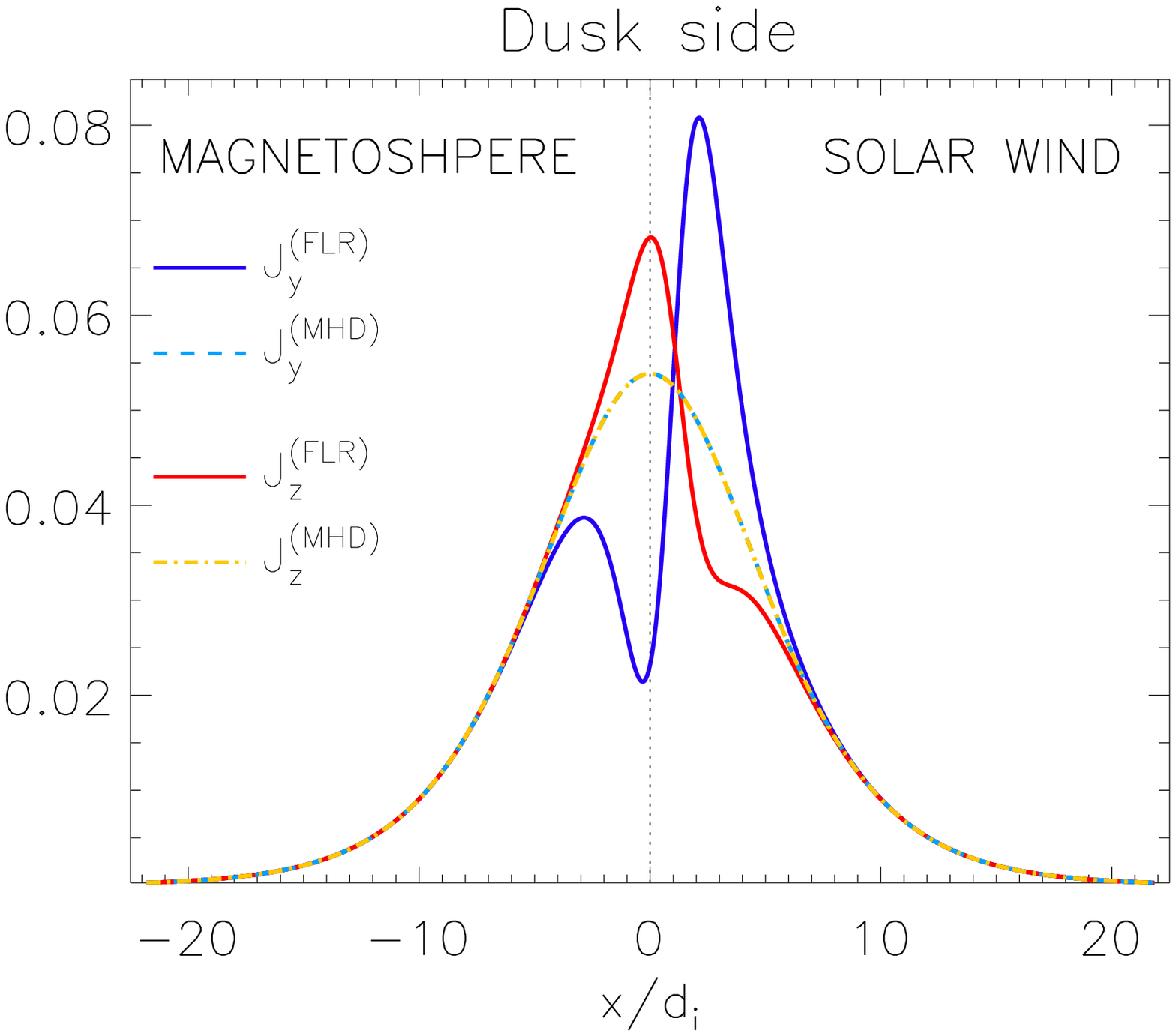}
 \caption{Current profiles for cases reported in Table~\ref{tab:tab1}. Top: case A, dawn (top left panel) and dusk (top right panel) sides. Bottom: case B, dawn (bottom left panel) and dusk (bottom right panel) sides. The MHD current profiles, $J_y^{\rm(MHD)}$ and $J_z^{\rm(MHD)}$, are reported with dashed light blue and dot-dashed orange lines, respectively, whereas the corresponding FLR-corrected profiles, $J_y^{\rm(FLR)}$ and $J_z^{\rm(FLR)}$, are drawn in blue and red solid lines, respectively.}
 \label{fig:Jprofiles_FLR}
\end{figure} 
%================================

Let us now consider few explicit examples relevant for the magnetopause layer and  see how the first-order FLR corrections qualitatively modify its current profile. 
For the sake of simplicity, we consider the case of $\vartheta=0$ and $\phi=\pi/2$ and two slightly different regimes are taken into account. 
A summary of the parameters adopted for the example profiles is given in Table~\ref{tab:tab1}. 
These parameter are chosen so that they are as realistic as possible for the low-latitude flanks of the magnetopause~\citep{HaalandJGRA2014}, as well as they are able to slightly emphasize some of the resulting features~\footnote{For instance, the choice of $L_B=6d_i$ is consistent with the mean thickness reported by \citet{HaalandJGRA2014} of $\simeq18\rho_i$ of the dawn side, whereas there is no explicit indication for the thickness of the velocity shear. In the present work, we have considered a velocity shear layer that is thinner that the magnetic shear layer and that are slightly shifted with respect to each other, in agreement with some other {\it Cluster} observations~\citep[e.g.,][]{FoullonJGRA2008,RossiPhDthesis2015}.}. 

\begin{figure}
\centering
 \includegraphics[width=0.496\textwidth]{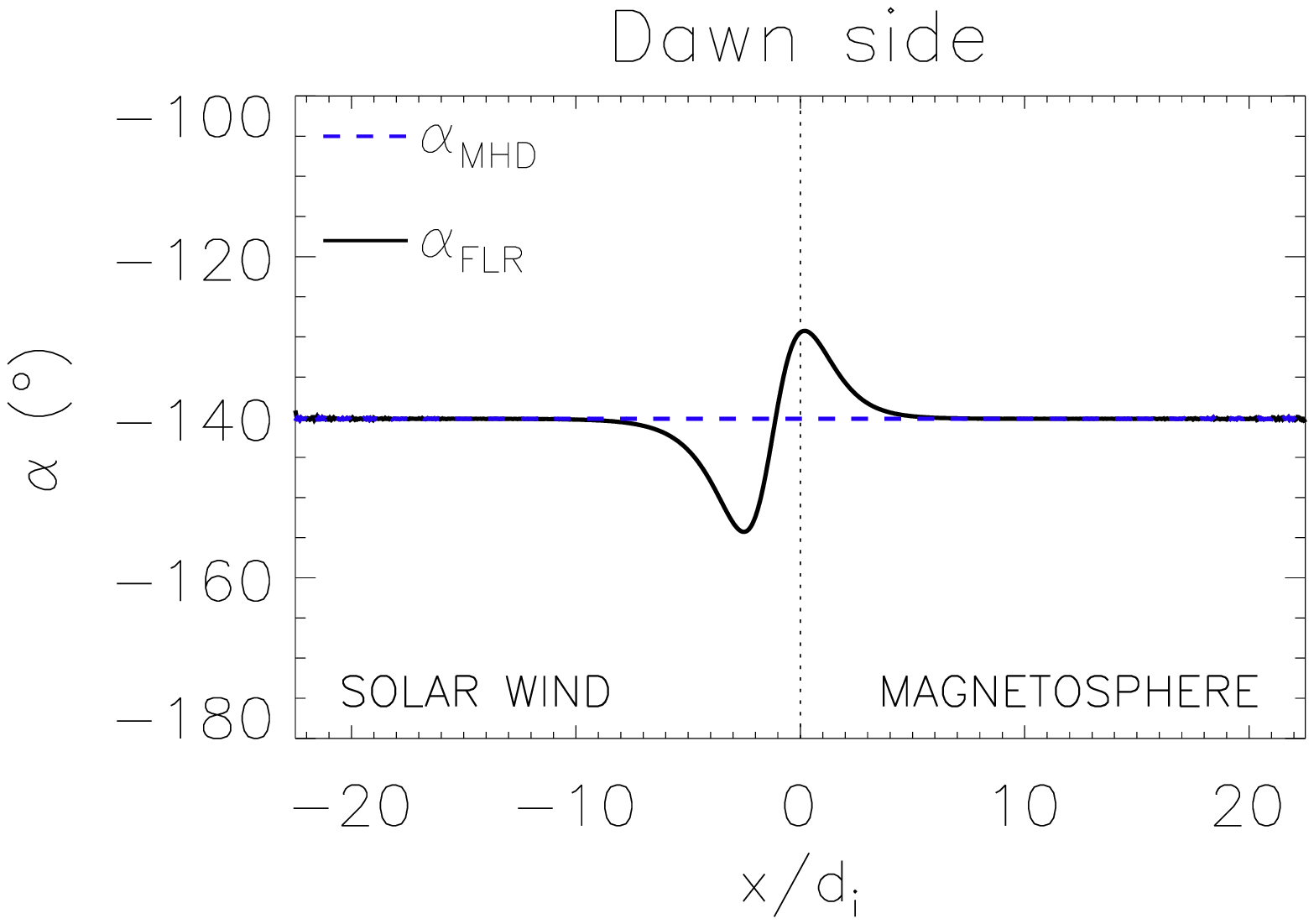}
 \includegraphics[width=0.496\textwidth]{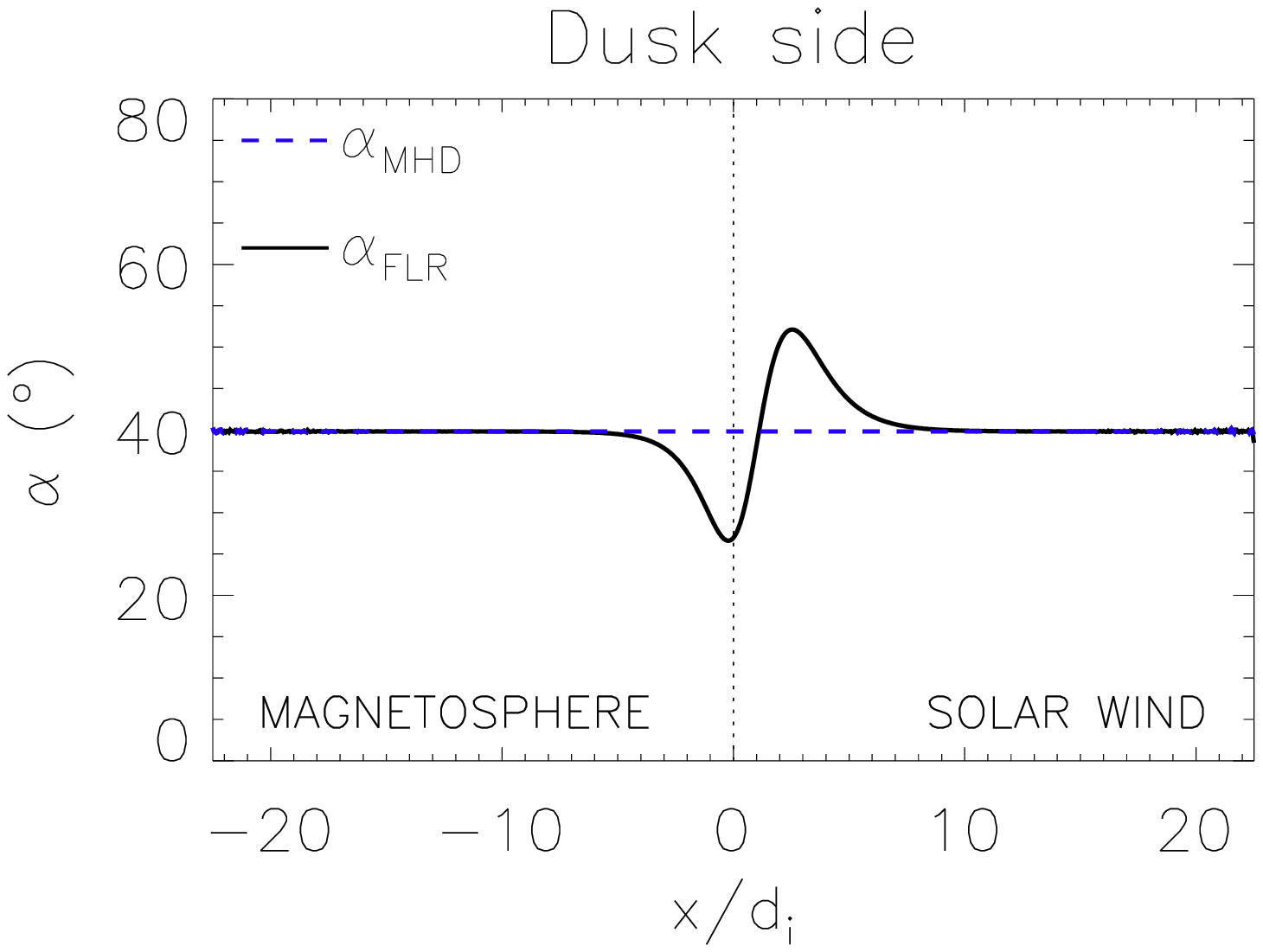}
 \includegraphics[width=0.496\textwidth]{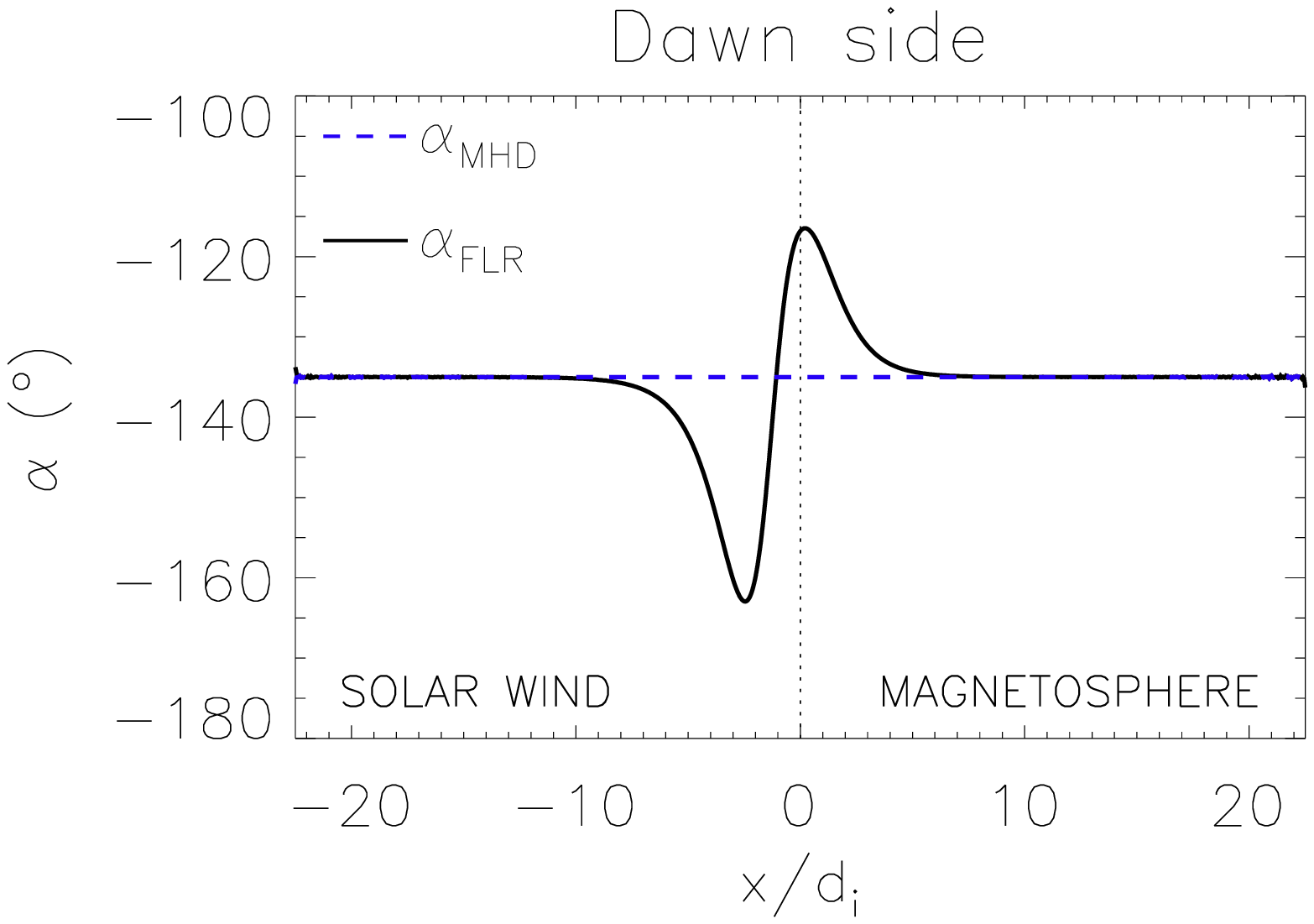}
 \includegraphics[width=0.496\textwidth]{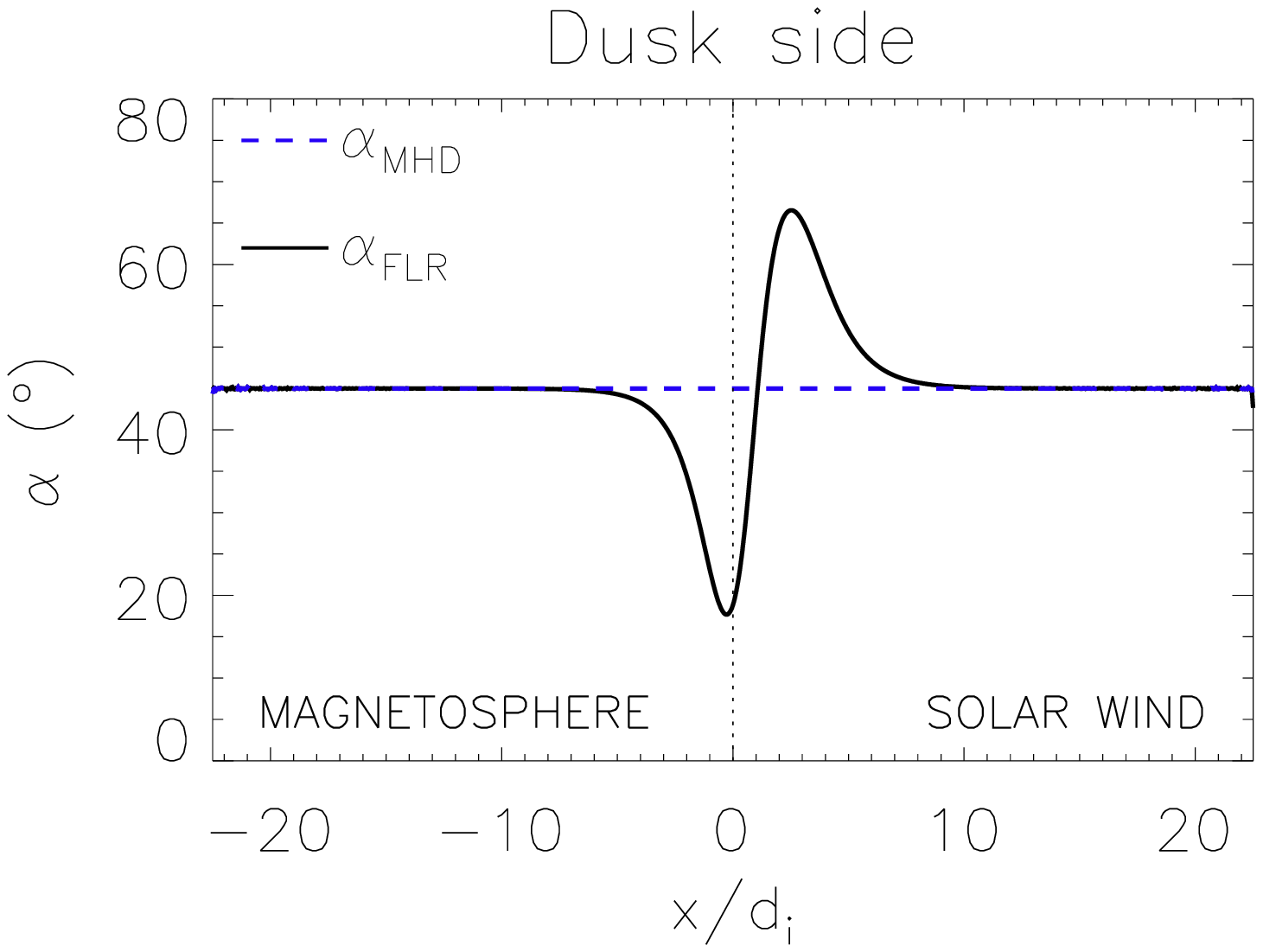}
 \caption{Profiles of the angle between the current $\Jv$ and the $z$-axis, $\alpha$, versus $x$ for cases reported in Fig.~\ref{fig:Jprofiles_FLR}. Top: case A, dawn (top left panel) and dusk (top right panel) sides. Bottom: case B, dawn (bottom left panel) and dusk (bottom right panel) sides.}
 \label{fig:ALPHAprofiles_FLR}
\end{figure}

In Fig.~\ref{fig:Jprofiles_FLR} we report the current profile arising from a simple MHD configuration, $J_y^{\rm(MHD)}$ and $J_z^{\rm(MHD)}$ (light blue dashed line and orange dot-dashed line, respectively), as well as the profile accounting for the first-order FLR corrections in Eqs.~(\ref{eq:Utildeprimo_KHequil}), $J_y^{\rm(FLR)}$ and $J_z^{\rm(FLR)}$ (blue and red solid lines, respectively).
The MHD profiles of the dusk and of the dawn sides, apart from the sign, have the same shape, i.e. it is the classic Chapman--Ferraro current layer~\citep{ChapmanFerraro1930}. 
On the other hand, the corresponding FLR-corrected profiles of the dawn and of the dusk sides are qualitatively different. This is the effect of the ``$\omegav\bv$ asymmetry'' intrinsically encoded in the FLR contributions.
Furthermore, the current structure of the shear layer in this latter case is much more complex than the Chapman--Ferraro MHD layer. 
In fact, a double-peak feature asymmetrically arises in $J^{\rm(FLR)}$ on the two sides of the flank magnetopause and the different modification of the two components of the current results in adjacent current sheets with different current direction (see Fig.~\ref{fig:ALPHAprofiles_FLR}, where we report the $x$-dependence of the angle between $\Jv$ and the $z$-axis, $\alpha=\arctan(J_y/J_z)$, for the cases shown in Fig.~\ref{fig:Jprofiles_FLR}).
These three peculiar features, namely (i) the dusk-dawn asymmetry of the current layer, (ii) the double-peak feature in the current profiles, and (iii) two (or more) adjacent current sheets having thickness of several ion Larmor radii and with different current directions, are qualitatively consistent with the Cluster observations reported in \citet{HaalandJGRA2014}. 
Taking into account these FLR effects can also be a relevant starting point for explaining certain anomalies occurring during magnetopause distortions related to large-scale magnetosheath plasma jets~\citep[see, e.g.,][]{DmitrievSuvorovaJGR2012}.

%================================
%================================

Finally, we want to stress that here we focused on the FLR corrections to the magnetic and current structures, as most of the analysis done on satellite data for the purpose of reconstructing the characteristics of the Earth's flank magnetopause has been carried out in this direction. However, there are other relevant features and signatures of non-ideal effects that one could seek for in the available satellite data, as, for instance, the equilibrium profiles presented here would be supported by agyrotropic particle distribution functions localized in the large-scale shear-flow layer at the Earth's magnetopause~\footnote{Clearly, here we are not taking into account additional deviations from isotropy (and from pure gyrotropy) due to local current and vorticity sheets forming in a turbulent plasma~\citep[see, e.g.,][]{ServidioPRL2012,ValentiniPOP2014,ValentiniNJP2016,FranciAIPC2016,CerriAPJL2018,Pezzi2018} and/or during reconnection events~\citep[see, e.g.,][]{ScudderDaughtonJGRA2008,AnuaiPOP2013}}.

\section{Conclusions}\label{sec:conclusions}

We have derived the one-dimensional equilibrium solutions for a shear-flow boundary layer within a so-called ``extended two-fluid'' (eTF) model accounting for first-order ion finite-Larmor-radius (FLR) corrections in the double-adiabatic limit.
These analytical solutions represent a generalization of the solutions given in \citealt{CerriPOP2013}.

We have explicitly shown that first-order FLR corrections exhibit what we have called ``$\omegav\bv$ asymmetry'', i.e., an asymmetry that depends on the relative orientation of the fluid vorticity, $\omegav$, and of the magnetic-field direction, $\bv$, through the scalar product $\omegav\boldsymbol{\cdot}\bv$.
Moreover, depending again on the parameter $\omegav\boldsymbol{\cdot}\bv$, it has been demonstrated that the free energy available in the shear flow is able to develop and sustain a non-negligible level of agyrotropy, i.e., a pressure (and temperature) anisotropy that is not limited to the directions parallel and perpendicular to the magnetic field (the so-called gyrotropy), but that manifests also within the plane perpendicular to $\bv$ as $p_\|\neq p_{\perp,1}\neq p_{\perp,2}$.

Finally, we have applied these FLR-corrected equilibrium profiles to few cases with parameters typical of the low-latitude flanks of the Earth's magnetopause. 
The resulting current structure has been shown to be more complex than the MHD layer by \citet{ChapmanFerraro1930}, in qualitative agreement with the {\it Cluster} observations recently reported in \citet{HaalandJGRA2014}. 
In particular, by accounting for ion FLR effects, we have been able to qualitatively reproduce the following key observational features: (i) an asymmetry of the current layer with respect to the dusk and the dawn sides of the magnetopause, (ii) a double-peak feature arising in the current profiles, and (iii) the presence of adjacent current sheets having thickness of several ion Larmor radii and with different current directions.
We want to stress that other effects that may contribute to further corrections have been neglected, e.g., the full ion pressure tensor dynamics and the electron kinetic effects, so a quantitative comparison between the {\it Cluster} data and our profiles would be beyond the scope of the present work. 
Nevertheless, the good qualitative agreement between our one-dimensional analytical profiles and the {\it Cluster} observations reported in \citet{HaalandJGRA2014} shows that ion FLR corrections are a relevant ingredient to correctly describe the Earth's flank magnetopause layer. 
Further effects, including a three-dimensional treatment of the magnetosphere-wind interface, as well as the full ion pressure tensor and self-consistent electron kinetic effects, will clearly have to be considered for a more quantitative comparison. 
In this regard, new and future space missions will also provide better measurements of the Earth's magnetopause structure and allow for a deeper understanding of the relevant plasma physics at play. %in that context.

Finally, we underline that the main consequences of the ion-FLR effects reported in this work, and their relation to anisotropy, agyrotropy, vorticity and current sheets, may have implications for a wide variety of astrophysical and space collisionless plasmas, from the turbulent solar wind to low-luminosity accretion flows around compact objects.

\acknowledgments

The author acknowledges F.~Pegoraro, F.~Califano, D.~Del~Sarto and A.~Tenerani for many valuable discussions on the subject in the past years, as well as M.~W.~Kunz for providing comments on the manuscript and the anonymous referees for comments that improved the clarity of the manuscript.
This work was completed while S.~S.~C. was supported by the National Aeronautics and Space Administration under Grant No. NNX16AK09G issued through the Heliophysics Supporting Research Program.

\appendix

\section{Derivation of the equilibrium configurations of a shear-flow layer with FLR effects}\label{app:generic_equil_derivation}

We now consider the case of a velocity-shear layer separating, for instance, two different plasmas. 
For the sake of simplicity, here we consider the one-dimensional equilibrium problem, which can be seen as a local approximation of the LLBL.
A class of analytical solutions to the 1D case that generalize the results provided in \citet{CerriPOP2013} and that include a much wider range of configurations of interest for what concerns magnetospheric observations will be provided.

\subsection{Preliminaries and assumptions}\label{subsec:assumptions}

In the following, we consider a given $x$-dependent incompressible MHD flow in the $y$-$z$ plane, 
%%%%%%%%%%%%%%%%%%%%%%%%%%%%%%%%%%%%%%%%%%%%%%%%%%%%%
\begin{equation}\label{app:eq:vel_shear_generic}
\uv=u_y(x)\ev_y+u_z(x)\ev_z\,,\qquad\bnabla\boldsymbol{\cdot}\uv\,=\,0\,,
\end{equation}
%%%%%%%%%%%%%%%%%%%%%%%%%%%%%%%%%%%%%%%%%%%%%%%%%%%%%
such that it becomes constant at the boundaries,
%%%%%%%%%%%%%%%%%%%%%%%%%%%%%%%%%%%%%%%%%%%%%%%%%%%%%
\begin{equation}\label{app:eq:U_bound}
  \lim_{x\to\pm\infty}u_y(x)\,=\,u_{0y}^{(\pm)}\,,\qquad
  \lim_{x\to\pm\infty}u_z(x)\,=\,u_{0z}^{(\pm)}\,,
\end{equation}
%%%%%%%%%%%%%%%%%%%%%%%%%%%%%%%%%%%%%%%%%%%%%%%%%%%%%
i.e., we consider a localized velocity shear (the vorticity is vanishing at the boundaries, $\lim_{x\to\pm\infty}\bnabla\boldsymbol{\times}\uv=0$).
The magnetic field also lies on the $y$-$z$ plane,
%%%%%%%%%%%%%%%%%%%%%%%%%%%%%%%%%%%%%%%%%%%%%%%%%%%%%
\[
\Bv(x)=B_{0y}H_y(x)\ev_y+B_{0z}H_z(x)\ev_z\,.
\]
%%%%%%%%%%%%%%%%%%%%%%%%%%%%%%%%%%%%%%%%%%%%%%%%%%%%%
The associated magnetic pressure is 
%%%%%%%%%%%%%%%%%%%%%%%%%%%%%%%%%%%%%%%%%%%%%%%%%%%%%
\begin{equation}\label{app:eq:P_B}
P_B(x)\,=\,\frac{B_0^2}{8\pi}\HH(x)\,,\quad
\HH(x)\,\equiv\,\frac{B_{0y}^2}{B_0^2}H_y^2(x)\,+\,\frac{B_{0z}^2}{B_0^2}H_z^2(x)\geq0\quad\forall x\,,
\end{equation}
%%%%%%%%%%%%%%%%%%%%%%%%%%%%%%%%%%%%%%%%%%%%%%%%%%%%%
where we have defined $B_0$ as the (constant) value of $|\Bv|$ at the right boundary ($x\to+\infty$):
%%%%%%%%%%%%%%%%%%%%%%%%%%%%%%%%%%%%%%%%%%%%%%%%%%%%%
\begin{equation}\label{app:eq:B0_def}
  B_0\,\equiv\,\lim_{x\to+\infty}\sqrt{B_{0y}^2H_y^2(x)+B_{0z}^2H_z^2(x)}\,,\quad
  \lim_{x\to+\infty}\HH(x)\,=\,1\,.
\end{equation}
%%%%%%%%%%%%%%%%%%%%%%%%%%%%%%%%%%%%%%%%%%%%%%%%%%%%%
We further assume a polytropic relation for the thermal pressures\footnote{Note that in the case considered here of incompressible flow, no heat fluxes and no gradients parallel to the magnetic field, the double-adiabatic relations and the dynamical pressure equations in the eTF model are equivalent to two different polytropic relations for $p_\|$ and $p_\perp$~\citep[see, e.g.,][]{CerriMSc2012,CerriPOP2014b,DelSartoPegoraroMNRAS2018}}: 
%%%%%%%%%%%%%%%%%%%%%%%%%%%%%%%%%%%%%%%%%%%%%%%%%%%%%
\begin{equation}\label{app:eq:pperp_generic_poly}
 \pperpp=\pperppz\,\FF_\perp(x)=\pperppz\left(\frac{n(x)}{n_0}\right)^{\gammaperpp}\,,\qquad
 \pperpe=\pperpez\,\GG_\perp(x)=\pperpez\left(\frac{n(x)}{n_0}\right)^{\gammaperpe}\,,
\end{equation}
%%%%%%%%%%%%%%%%%%%%%%%%%%%%%%%%%%%%%%%%%%%%%%%%%%%%%
and
%%%%%%%%%%%%%%%%%%%%%%%%%%%%%%%%%%%%%%%%%%%%%%%%%%%%%
\begin{equation}\label{app:eq:ppara_generic_poly}
 \pparap=\pparapz\,\FF_\|(x)=\pparapz\left(\frac{n(x)}{n_0}\right)^{\gammaparap}\,,\qquad
 \pparae=\pparaez\,\GG_\|(x)=\pparaez\left(\frac{n(x)}{n_0}\right)^{\gammaparae}\,,
\end{equation}
%%%%%%%%%%%%%%%%%%%%%%%%%%%%%%%%%%%%%%%%%%%%%%%%%%%%%
where $\FF_\perp$, $\FF_\|$, $\GG_\perp$, and $\GG_\|$ are functions that reduce to unity for $x\to+\infty$, as it is for $\HH$.

\subsection{General one-dimensional MHD solutions for incompressible flows}\label{app:subsec:1D_MHDgeneric}

Within an (anisotropic) MHD model of plasma, the shear-flow does not play a role in the equilibrium profile. 
In fact, when $\bpip^{(1)}$ is neglected, the equilibrium condition for the above configuration simply consists of a balance between the magnetic pressure, $B^2(x)/8\pi$, and total perpendicular thermal pressures, $P_\perp(x)$: 
%%%%%%%%%%%%%%%%%%%%%%%%%%%%%%%%%%%%%%%%%%%%%%%%%%%%%
\begin{equation}\label{app:eq:equil-cond_HMHDaniso}
  \frac{\rm d}{{\rm d}x}\left[\pperpp(x)+\pperpe(x)+\frac{B^2(x)}{2}\right]\,=\,0\,.
\end{equation}
%%%%%%%%%%%%%%%%%%%%%%%%%%%%%%%%%%%%%%%%%%%%%%%%%%%%%
In particular, the above condition allows also the widely adopted uniform and homogeneous plasma configuration: 
$\pperpa=\pparaaz$, $\pperpa=\pperpaz$, $B_y=B_{0y}$, and $B_z=B_{0z}$. 
Such homogeneous profiles are not an equilibrium solution when FLR corrections (or the full pressure-tensor equations) are included in the fluid description~\citep{CerriMSc2012,CerriPOP2013,CerriPOP2014b}, unless the velocity profile is a linear function of $x$ (see \S~\ref{app:subsec:FLRequil_generic}). 
In general, the solution of the MHD equilibrium condition in (\ref{app:eq:equil-cond_HMHDaniso}) is completely described by the magnetic pressure profile in (\ref{app:eq:P_B}), which determines all the other relevant functions, $\FF_\perp(x)$ and $\GG_\perp(x)$. 
In fact, assuming $\gammaperpe=\gammaperpp$ for simpicity, quasi-neutrality reads as 
%%%%%%%%%%%%%%%%%%%%%%%%%%%%%%%%%%%%%%%%%%%%%%%%%%%%%
\begin{equation}\label{app:eq:GG-FF_implicit}
\GG_\perp(x)\,=\,\FF_\perp(x)
\end{equation}
%%%%%%%%%%%%%%%%%%%%%%%%%%%%%%%%%%%%%%%%%%%%%%%%%%%%% 
and the equilibrium condition finally gives $\FF_\perp$ as function of $\HH$,
%%%%%%%%%%%%%%%%%%%%%%%%%%%%%%%%%%%%%%%%%%%%%%%%%%%%%
\begin{equation}\label{app:eq:HH-FF_implicit}
\FF_\perp(x)\,=\,1\,+\,\frac{1}{\betaperpz}\Big[1-\HH(x)\Big]\,,
\end{equation}
%%%%%%%%%%%%%%%%%%%%%%%%%%%%%%%%%%%%%%%%%%%%%%%%%%%%%
where $\betaperpz=\betaperppz+\betaperpez$ (with $\betaperpaz\equiv8\pi\,\pperpaz/B_0^2$), and the constant is set to $1+\betaperpz$ by the boundary conditions at $x\to+\infty$ 
(the requirement $\FF_\perp(x)\to1$ for $x\to+\infty$ is then automatically satisfied due to (\ref{app:eq:B0_def})). 
Furthermore, since the function $\FF_\perp(x)$ is related to the thermal pressure, it cannot assume negative values, which provides the additional condition
%%%%%%%%%%%%%%%%%%%%%%%%%%%%%%%%%%%%%%%%%%%%%%%%%%%%%
\begin{equation}\label{app:eq:HH-FF_implicit_2}
\FF_\perp(x)\,\geq\,0\quad\forall x\,\quad
\Longleftrightarrow\quad\,
\HH(x)\,\leq\,1\,+\,\betaperpz\,.
\end{equation}
%%%%%%%%%%%%%%%%%%%%%%%%%%%%%%%%%%%%%%%%%%%%%%%%%%%%%
This states physically that any variation of the magnetic pressure, $\Delta B^2/8\pi=(B^2(x)-B_0^2)/8\pi$, cannot exceed the total thermal pressure, $P_{\perp,0}=\pperppz+\pperpez$, where $B_0$, $\pperppz$ and $\pperpez$ are the values at $x\to+\infty$.
The parallel thermal pressures follow from the polytropic assumption, e.g., $\FF_\|(x)=[\FF_\perp(x)]^{\gammaparap/\gammaperpp}$. 
Analogously, the temperature profiles follow from $T_{\perp\alpha}=p_{\perp\alpha}/n$ and $T_{\|\alpha}=p_{\|\alpha}/n$.

Starting from this MHD class of solutions, we self-consistently derive the corresponding equilibrium profiles with first-order FLR corrections.

\subsection{General first-order FLR corrections to the one-dimensional MHD solutions}\label{app:subsec:FLRequil_generic}

Let us now consider the changes to the MHD equilibrium profiles derived above that are induced 
by the velocity shear in (\ref{app:eq:vel_shear_generic}) when first-order FLR corrections are taken into account. 
In this case, the only component of $\bpip^{(1)}$ that is relevant to the equilibrium condition is
%%%%%%%%%%%%%%%%%%%%%%%%%%%%%%%%%%%%%%%%%%%%%%%
\begin{equation}\label{app:eq:pi_xx_equil_generic}
        \pi_{{\rm p},xx}^{(1)}\, =\,  
        -\,\frac{1}{2}\,\frac{\mpp c}{e|\Bv|}
        \left(b_z\frac{{\rm d}\, u_y}{{\rm d}x}\,-\,b_y\frac{{\rm d}\, u_z}{{\rm d}x}\right)\pperpp\,.
\end{equation}
%%%%%%%%%%%%%%%%%%%%%%%%%%%%%%%%%%%%%%%%%%%%%%%%%%%%%
The equilibrium condition in (\ref{app:eq:equil-cond_HMHDaniso}) now reads
%%%%%%%%%%%%%%%%%%%%%%%%%%%%%%%%%%%%%%%%%%%%%%%%%%%%%
\begin{equation}\label{app:eq:equil-cond_eHMHD}
  \frac{\rm d}{{\rm d}x}\left\{
  \left[1-\frac{1}{2}\frac{B_z(x)u_y'(x)-B_y(x)u_z'(x)}{e\,B^2(x)/\mpp c}\right]\pperpp(x)\,+\,\pperpe(x)\,+\,\frac{B^2(x)}{8\pi}\right\}\,=\,0\,,
\end{equation}
%%%%%%%%%%%%%%%%%%%%%%%%%%%%%%%%%%%%%%%%%%%%%%%%%%%%%
where the prime denotes the $x$-derivative.  
The above expressions can be explicitly written in terms of the fluid vorticity, $\boldsymbol{\omega}\equiv\bnabla\boldsymbol{\times}\uv$, and of the magnetic field direction, $\bv$:  
%%%%%%%%%%%%%%%%%%%%%%%%%%%%%%%%%%%%%%%%%%%%%%%%%%%%%
\begin{equation}\label{app:eq:pi_xx_equil-cond_generic_vort}
  \pi_{{\rm p},xx}^{(1)}\, =\,  
    -\,\frac{1}{2}\frac{\mpp c}{e B}(\bv\boldsymbol{\cdot}\boldsymbol{\omega})\,\pperpp\quad\longrightarrow\quad
   \frac{\rm d}{{\rm d}x}\left[
    \left(1-\frac{\mpp c}{eB}\frac{\bv\boldsymbol{\cdot}\boldsymbol{\omega}}{2}\right)\pperpp\,+\,\pperpe\,+\,\frac{B^2}{8\pi}\right]\,=\,0\,,
\end{equation}
%%%%%%%%%%%%%%%%%%%%%%%%%%%%%%%%%%%%%%%%%%%%%%%%%%%%%
where $\omega_y=-u_z'$ and $\omega_z=u_y'$ are the components of the fluid vorticity in our configuration.
The dependence on $\bv\boldsymbol{\cdot}\boldsymbol{\omega}$ highlights the intrinsic asymmetry in the system due to FLR corrections and related to the degree of alignment (or anti-alignment) between the flow vorticity and the magnetic field. 
We stress, however, that the simple dependence on the vorticity and magnetic-field direction in (\ref{app:eq:pi_xx_equil_generic}) is due to the 1D character of the problem considered here.

We now seek FLR-corrected equilibrium profiles in the form $\tFF_\perp(x)=\FF_\perp(x)f_\perp(x)$, $\tGG_\perp(x)=\GG_\perp(x)g_\perp(x)$ and $\tHH(x)=\HH(x)h(x)$, where $f_\perp$, $g_\perp$ and $h$ 
are the ``correction functions''. 
Due to the boundary conditions on the MHD flow, (\ref{app:eq:U_bound}), the gyroviscous tensor vanishes at the boundaries, $\lim_{x\to\pm\infty}\bpip^{(1)}=0$, and thus the correction functions must reduce to unity accordingly, $\lim_{x\to\pm\infty}\{f_\perp(x),g_\perp(x),h(x)\}=1$.
Therefore, $\tFF_\perp$, $\tGG_\perp$ and $\tHH$ reduce to the corresponding MHD profiles away from the shear layer, where the vorticity vanishes (or, in general, where the vorticity becomes uniform and homogeneous). 
Moreover, since we want to preserve quasi-neutrality, $\tFF(x)=\tGG(x)$ must hold and therefore, 
using (\ref{app:eq:GG-FF_implicit}), we obtain the condition 
%%%%%%%%%%%%%%%%%%%%%%%%%%%%%%%%%%%%%%%%%%%%%%%%%%%%%
\begin{equation}\label{app:eq:g-f_implicit}
g_\perp(x)\,=\,f_\perp(x)\,.
\end{equation}
%%%%%%%%%%%%%%%%%%%%%%%%%%%%%%%%%%%%%%%%%%%%%%%%%%%%%
In order to relate $h(x)$ and $f_\perp(x)$, we actually need to impose a further constraint on the equilibrium.
Such a condition cannot be derived from first principles and would rather be driven by a physical interpretation of the problem under study. 
Here we provide a viable option based on the plasma beta 
parameter~\citep[see, e.g.,][for examples about different constraints]{CerriPOP2013,CerriPOP2014b}.
Since the (thermal) Larmor radius is sensitive to the (perpendicular) plasma beta, 
a very reasonable constraint is to require that the MHD profile $\betaperpp(x)$ 
does not change when passing to the corresponding FLR-corrected profile, i.e.,
%%%%%%%%%%%%%%%%%%%%%%%%%%%%%%%%%%%%%%%%%%%%%%%%%%%%%
\begin{equation}\label{app:eq:h-f_implicit}
\betaperpp(x)\Big|_{\rm MHD}\,=\,\betaperpp(x)\Big|_{\rm MHD+FLR}\,\quad\Longrightarrow
\quad\,h(x)=f_\perp(x)\,.
\end{equation}
%%%%%%%%%%%%%%%%%%%%%%%%%%%%%%%%%%%%%%%%%%%%%%%%%%%%%
Then, using the above relations and the boundary conditions at $x\to+\infty$ to set 
the integration constant to $1+\betaperpz$, from (\ref{app:eq:equil-cond_eHMHD}) we obtain the following equation for $f_\perp(x)$:
%%%%%%%%%%%%%%%%%%%%%%%%%%%%%%%%%%%%%%%%%%%%%%%%%%%%%
\begin{equation}\label{app:eq:sqrt-f_Blocal}
 f_\perp(x)\,-\,\tUp(x)\,\sqrt{f_\perp(x)}\,-\,1\,=\,0\,,
\end{equation}
%%%%%%%%%%%%%%%%%%%%%%%%%%%%%%%%%%%%%%%%%%%%%%%%%%%%%
where we have defined
%%%%%%%%%%%%%%%%%%%%%%%%%%%%%%%%%%%%%%%%%%%%%%%%%%%%%
\begin{equation}\label{app:eq:Utilde_def}
 \tUp(x)\,\equiv\,\frac{\tbetaperppz}{2}\,\frac{\mpp c}{e\,B_0}\,
 \frac{\FF_\perp(x)}{\HH(x)}\left(\frac{B_{0z}}{B_0}H_z(x)u_y'(x)\,
 -\,\frac{B_{0y}}{B_0}H_y(x)u_z'(x)\right)
\end{equation}
%%%%%%%%%%%%%%%%%%%%%%%%%%%%%%%%%%%%%%%%%%%%%%%%%%%%%
with $\tbetaperppz\equiv\betaperppz/(1+\betaperpz)$ for brevity. 
Note that the above equation for $f_\perp(x)$ has been obtained taking into account the FLR corrections computed with the self-consistent equilibrium magnetic field profile, $B(x)=B_0\sqrt{\HH(x)f_\perp(x)}$ (we remind that $h(x)=f_\perp(x)$ holds). 
Finally, since $\pperpp(x)$ must be a positive quantity, we require $f_\perp(x)\geq0$ $\forall\,x$, so that the only physical solution of (\ref{app:eq:sqrt-f_Blocal}) is 
%%%%%%%%%%%%%%%%%%%%%%%%%%%%%%%%%%%%%%%%%%%%%%%%%%%%%
\begin{equation}\label{app:eq:f_Blocal_solution}
 f_\perp(x)\,=\,\left\{\frac{\tUp(x)}{2}\,+\sqrt{1+\left(\frac{\tUp(x)}{2}\right)^2\,}\,\right\}^2\,\,.
\end{equation}
%%%%%%%%%%%%%%%%%%%%%%%%%%%%%%%%%%%%%%%%%%%%%%%%%%%%%
This correctly reduces to unity for vanishing FLR terms, $\tUp\to0$, recovering the MHD profiles. 
The resulting FLR-corrected profiles are therefore given by
%%%%%%%%%%%%%%%%%%%%%%%%%%%%%%%%%%%%%%%%%%%%%%%%%%%%%
\begin{equation}\label{app:eq:pperpp-pparap_FLRcorrected}
 \pperpp(x)\,=\,\pperppz\,\FF_\perp(x)\,f_\perp(x)\,,\qquad
 \pparap(x)\,=\,\pparapz\Big(\FF_\perp(x)\,f_\perp(x)\Big)^{\gammapara/\gammaperp}\,,
\end{equation}
\begin{equation}\label{app:eq:den_FLRcorrected}
 n(x)\,=\,n_0\Big(\FF_\perp(x)\,f_\perp(x)\Big)^{1/\gammaperp}\,,
\end{equation}
%%%%%%%%%%%%%%%%%%%%%%%%%%%%%%%%%%%%%%%%%%%%%%%%%%%%%
%%%%%%%%%%%%%%%%%%%%%%%%%%%%%%%%%%%%%%%%%%%%%%%%%%%%%
\begin{equation}\label{app:eq:B_FLRcorrected}
 B_y(x)\,=\,B_{0y}\,H_y(x)\sqrt{f_\perp(x)}\,,\qquad
 B_z(x)\,=\,B_{0z}\,H_z(x)\sqrt{f_\perp(x)}\,,
\end{equation}
%%%%%%%%%%%%%%%%%%%%%%%%%%%%%%%%%%%%%%%%%%%%%%%%%%%%%
from which the current density, $\Jv=\nabla\times\Bv$, follows.

\section{Derivation of the first-order FLR contributions: a perturbative approach}\label{app:1st-order-FLR}

In this Appendix, we provide a derivation of the finite Larmor radius corrections to the gyrotropic pressure tensor based on a perturbative expansion of the full pressure tensor dynamic equation\footnote{For a derivation based on a perturbative expansion of the distribution function, see \cite{MacmahonPOF1965} or \cite{SchekochihinMNRAS2010}. Other classical derivations can be found in \citet{YajimaPTP1966}, \citet{RamosPOP2005b} or in \citet{MjolhusNPG2009}.}. 
Further, we explicitly comment on the symmetry properties of the perturbed equations and the correspondent solutions, which has a direct relevance for many configurations with a velocity shear.

Note that in the remainder of this Appendix we are going to drop the species index everywhere, except when it is needed (e.g., when the sign of the charge matters).

\subsection{Perturbative expansion of the pressure tensor equation}

Let us consider the dynamic equation for the full pressure tensor,
%%%%%%%%%%%%%%%%%%%%%%%%%%%%%%%%%%%%%%%%%%%%%%%%%%%%%
\begin{equation}\label{eq:app:press-tens_1-b}
  \frac{\partial\,\Pi_{ij}}{\partial t}\, +\, \frac{\partial}{\partial x_k}\Big(\Pi_{ij}u_k + Q_{ijk}\Big)\, +\,
  \Pi_{ik}\frac{\partial\,u_j}{\partial x_k}\, +\, \Pi_{jk}\frac{\partial\,u_i}{\partial x_k}\, =\
  \Omegaca\Big(\epsilon_{ikl}\Pi_{jk} + \epsilon_{jkl}\Pi_{ik}\Big)b_l\,,
 \end{equation}
%%%%%%%%%%%%%%%%%%%%%%%%%%%%%%%%%%%%%%%%%%%%%%%%%%%%%
where $\eijk$ is the Levi--Civita symbol, and perturbatively expand it with respect to the small parameter
%%%%%%%%%%%%%%%%%%%%%%%%%%%%%%%%%%%%%%%%%%%%%%%%%%%%%
\[
 \varepsilon\,\equiv\,\frac{\rho}{L}\,\sim\,\frac{\omega}{\Omega}\,\ll1\,,
\]
%%%%%%%%%%%%%%%%%%%%%%%%%%%%%%%%%%%%%%%%%%%%%%%%%%%%%
where $\rho$ is the Larmor radius, $L$ is the typical length scale of variation of fluid quantities, and $\omega\sim u/L$ is the characteristic frequency of the fluid dynamics. 
Here we adopt the so-called ``fast-dynamics ordering'', $u\sim\vth$~\citep{MacmahonPOF1965,RamosPOP2005a,CerriPOP2013}. 
Using dimensionless quantities denoted by a tilde\footnote{We normalize all the quantities with respect to the mass, $m$, the thermal speed, $\vth$, and a reference density and magnetic field, $n_0$ and $B_0$, respectively: $n=n_0\widetilde{n}$, $B=B_0\widetilde{B}$, $u=\vth\widetilde{u}$, $\Pi=mn_0\vth^2\widetilde{\Pi}$, and $Q=mn_0\vth^3\widetilde{Q}$. The derivatives, are normalized as $\partial/\partial x=L^{-1}\partial/\partial\widetilde{x}$ and $\partial/\partial t=\tau^{-1}\partial/\partial\widetilde{t}$, with the ordering $L/\tau\sim u\sim\vth$.}, equation (\ref{eq:app:press-tens_1-b}) rewrites as
%%%%%%%%%%%%%%%%%%%%%%%%%%%%%%%%%%%%%%%%%%%%%%%%%%%%%%%
\[
 \Big(\epsilon_{ikl}\widetilde{\Pi}_{jk}\, +\, \epsilon_{jkl}\widetilde{\Pi}_{ik}\Big)\widetilde{b}_l\, =
\]
%%%%%%%%%%%%%%%%%%%%%%%%%%%%%%%%%%%%%%%%%%%%%%%%%%%%%%%
\begin{equation}\label{eq:app:press-tens_1-c}
 =\, \varepsilon\,\frac{\sigma_\alpha}{\big|\widetilde{\Bv}\big|}\left[
 \frac{\partial\,\widetilde{\Pi}_{ij}}{\partial\widetilde{t}}\, +\, 
 \frac{\partial}{\partial\widetilde{x}_k}\Big(\widetilde{\Pi}_{ij}\widetilde{u}_k\Big)\, +\,
 \widetilde{\Pi}_{ik}\frac{\partial\,\widetilde{u}_j}{\partial\widetilde{x}_k}\, +\, 
 \widetilde{\Pi}_{jk}\frac{\partial\,\widetilde{u}_i}{\partial\widetilde{x}_k}\, +\,
 \frac{\partial\,\widetilde{Q}_{ijk}}{\partial\widetilde{x}_k}\right]\,,
\end{equation}
%%%%%%%%%%%%%%%%%%%%%%%%%%%%%%%%%%%%%%%%%%%%%%%%%%%%%%%
where we have defined $\sigma_\alpha\equiv\textrm{sign}(e_\alpha)$, i.e., the sign embedded in the cyclotron frequency, $\Omegaca=e_\alpha B_0/\ma c=\sigma_\alpha|e_\alpha|B_0/\ma c\equiv\sigma_\alpha|\Omegaca|$. We then expand the pressure tensor and heat flux tensor in powers of $\varepsilon$, i.e.
%%%%%%%%%%%%%%%%%%%%%%%%%%%%%%%%%%%%%%%%%%%%%%%%%%%%%%%
\begin{equation}\label{eq:app:press-heat-expansions}
 \widetilde{\Pi}_{ij}\, =\, \sum_{n=0}^\infty \varepsilon^n\, \widetilde{\Pi}_{ij}^{(n)}\qquad\textrm{and}\qquad
 \widetilde{Q}_{ijk}\, =\, \sum_{n=0}^\infty \varepsilon^n\,  \widetilde{Q}_{ijk}^{(n)}\,.
\end{equation}
%%%%%%%%%%%%%%%%%%%%%%%%%%%%%%%%%%%%%%%%%%%%%%%%%%%%%%%
Hereafter, the tilde will be omitted for the sake of simplicity and all the quantities have to be understood as dimensionless. The $n$th-order pressure tensor equation then reads
%%%%%%%%%%%%%%%%%%%%%%%%%%%%%%%%%%%%%%%%%%%%%%%%%%%%%%%
\begin{equation}\label{eq:app:ptoperators}
 \mathcal{L}_\Bv\big[\Pi_{ij}^{(n)}\big]\, =\, \mathcal{R}_{\uv}\big[\Pi_{ij}^{(n-1)}\big]\,
 +\, \mathcal{D}\big[ Q_{ij(k)}^{(n-1)}\big]\,,
\end{equation}
%%%%%%%%%%%%%%%%%%%%%%%%%%%%%%%%%%%%%%%%%%%%%%%%%%%%%%%
where we have introduced the following linear opeartors:
%%%%%%%%%%%%%%%%%%%%%%%%%%%%%%%%%%%%%%%%%%%%%%%%%%%%%
\begin{eqnarray}\label{eq:app:lin-operators-def}
\mathcal{L}_\Bv\big[\bPi\big] & \equiv & \{\bPi\times\bv\}^{\rm(sym)}\\
\mathcal{R}_\uv\big[\bPi\big] & \equiv & \frac{{\rm d}\,\bPi}{{\rm d}t}\,+\,\bPi\big(\bnabla\cdot\uv\big)\, +\, \{\bPi:\bnabla\uv\}^{\rm(sym)}\\
\mathcal{D}\big[\bQ\big] & \equiv & \bnabla\cdot\bQ\,,
\end{eqnarray}
%%%%%%%%%%%%%%%%%%%%%%%%%%%%%%%%%%%%%%%%%%%%%%%%%%%%%
which contribute to the evolution of the pressure tensor by involving only $\Bv$, $\uv$ and $\bQ$, respectively ($\partial/\partial t + \uv\cdot\bnabla$ has been replaced by the Lagrangian time derivative ${\rm d}/{\rm d}t$ for shortness). The zero order, $n=0$, gives
%%%%%%%%%%%%%%%%%%%%%%%%%%%%%%%%%%%%%%%%%%%%%%%%%%%%%%%%%%%%
\begin{equation}\label{eq:app:PIzeroth-eq}
\Big(\epsilon_{ilm} \Pi_{lj}^{(0)}\ +\ \epsilon_{jlm} \Pi_{li}^{(0)}\Big)b_m\ =\ 0\,,
\end{equation}
%%%%%%%%%%%%%%%%%%%%%%%%%%%%%%%%%%%%%%%%%%%%%%%%%%%%%%%%%%%%
that means that $\Pi_{ij}^{(0)}$ belongs to the kernel of the $\mathcal{L}_\Bv$ operator, whereas the first-order equation, $n=1$, is 
%%%%%%%%%%%%%%%%%%%%%%%%%%%%%%%%%%%%%%%%%%%%%%%%%%%%%%%%%%%%
\begin{equation}\label{eq:app:PIfirst-eq}
\Big(\epsilon_{ilm} \Pi_{lj}^{(1)}\, +\, \epsilon_{jlm} \Pi_{li}^{(1)}\Big)b_m\,  =\, \frac{\sigma_\alpha}{B}\left[\frac{{\rm d}\Pi_{ij}^{(0)}}{{\rm d} t} + 
\Pi_{ij}^{(0)}\frac{\partial u_k}{\partial x_k} +
\Pi_{ik}^{(0)}\frac{\partial u_j}{\partial x_k} + 
\Pi_{jk}^{(0)}\frac{\partial u_i}{\partial x_k} +
\frac{\partial Q_{ijk}^{(0)}}{\partial x_k}\right]\,.
\end{equation}
%%%%%%%%%%%%%%%%%%%%%%%%%%%%%%%%%%%%%%%%%%%%%%%%%%%%%%%%%%%%
Before proceeding in the actual solution of the above equations, let us comment on their symmetry properties, in particular with respect to the magnetic field direction.

\subsection{Symmetry considerations on the perturbed equations}\label{app:subsec_symmetries}

Let us consider the three operators, ${\cal L}_\Bv$, ${\cal R}_\uv$ and ${\cal D}$. If we invert the direction of the magnetic field, $\mathbf{B \to-B}$, then such operators transform as
%%%%%%%%%%%%%%%%%%%%%%%%%%%%%%%%%%%%%%%%%%%%%%%%%%%%%
\begin{eqnarray}\label{eq:app:lin-operators-Binv}
\mathcal{L}_\Bv\big[\bullet\big]\, & \quad\to\quad & {\cal L}_{-\Bv}\big[\bullet\big]\,=\,-\,\mathcal{L}_\Bv\big[\bullet\big]\\
\mathcal{R}_\uv\big[\bullet\big]\, & \quad\to\quad & \quad\mathcal{R}_\uv\big[\bullet\big]\\
\mathcal{D}\big[\bullet\big]\, & \quad\to\quad & \quad\,\mathcal{D}\big[\bullet\big]\,,
\end{eqnarray}
%%%%%%%%%%%%%%%%%%%%%%%%%%%%%%%%%%%%%%%%%%%%%%%%%%%%%
and this symmetry property has a direct consequence on the solutions. 

Let us consider the zeroth-order equation, (\ref{eq:app:PIzeroth-eq}), and a possible solution $\Pi_+^{(0)}$. 
Then, if we reverse the direction of the magnetic field, the linear operator ${\cal L}_\Bv$ also changes sign, but the zeroth-order equation remains the same and $\bPi_+^{(0)}$ is still a solution (i.e., if $\bPi_-^{(0)}$ is the solution when the magnetic field direction is reversed, then $\bPi_-^{(0)}=\bPi_+^{(0)}$ must hold in order to have a unique solution). 
Therefore, $\bPi^{(0)}$ is invariant under magnetic field inversion and we can drop the ``$+$'' and ``$-$'' subscript (see \S~\ref{app:subsec:CGLsolution}). 

Let $\bPi_+^{(1)}$ be a solution of the first-order equation (\ref{eq:app:PIfirst-eq}),
%%%%%%%%%%%%%%%%%%%%%%%%%%%%%%%%%%%%%%%%%%%%%%%%%%%%%
\[
\mathcal{L}_\Bv\big[\bPi_+^{(1)}\big]\, =\, 
\mathcal{R}_\uv\big[\bPi^{(0)}\big]\, +\,
\mathcal{D}\big[\bQ^{(0)}\big]\,.
\]
%%%%%%%%%%%%%%%%%%%%%%%%%%%%%%%%%%%%%%%%%%%%%%%%%%%%%
Now consider the same configuration, but with just the magnetic field in the opposite direction, i.e. $\bv\to-\bv$. Regardless of the actual behavior of the gyrotropic heat-flux tensor, $\bQ^{(0)}$, with respect to such inversion\footnote{One can show that $\bQ^{(0)}$ has to be a solution of ${\cal L}_\Bv[\bQ^{(0)}]=0$ and it will therefore be a combination of the type $\bQ^{(0)}=q_\|\bv\bv\bv+q_\perp\{\btau\bv\}^{\rm(sym)}$~\citep{GoswamiPOP2005}. This means that the gyrotropic heat-flux tensor changes sign when $\bv\to-\bv$. However, this does not play a role in the following argument.}, if we assume that the first-order solution $\bPi_+^{(1)}$ is invariant with respect to $\bv\to-\bv$, we then obtain a different equation:
%%%%%%%%%%%%%%%%%%%%%%%%%%%%%%%%%%%%%%%%%%%%%%%%%%%%%
\[
\mathcal{L}_\Bv\big[\bPi_+^{(1)}\big]\, =\,-\, 
\mathcal{R}_\uv\big[\bPi^{(0)}\big]\, \mp\,
\mathcal{D}\big[\bQ^{(0)}\big]\,,
\]
%%%%%%%%%%%%%%%%%%%%%%%%%%%%%%%%%%%%%%%%%%%%%%%%%%%%%
where the $\mp$ sign in front of ${\cal D}[\bQ^{(0)}]$ takes into account for any possible behavior of $\bQ^{(0)}$ with respect to such inversion. 
Let us drop the heat-flux contribution for a moment and consider the two equations, ${\cal L}_\Bv[\bPi_+^{(1)}]={\cal R}_\uv[\bPi^{(0)}]$ and ${\cal L}_\Bv[\bPi_+^{(1)}]=-{\cal R}_\uv[\bPi^{(0)}]$. 
Clearly, a non-zero solution $\bPi_+^{(1)}$ cannot satisfy simultaneously the two equations above, and so we must admit that there exists a different solution, $\bPi_-^{(1)}$. 
Due to the linear nature of the operators, it is immediate to see that a relation $\bPi_-^{(1)}=-\bPi_+^{(1)}$ must hold. With the contribution of the heat flux the relation might not be straightforward as $\bPi_-^{(1)}=-\bPi_+^{(1)}$, but, again, being ${\cal L}_\Bv$, ${\cal R}_\uv$ and ${\cal D}$ linear operators, there will be anyway a part of $\bPi^{(1)}$ that changes sign when $\bv\to-\bv$. This is a feature deeply encoded in the governing equations of a plasma, but it first emerges only when the fluid hierarchy is retained up to the pressure tensor equation~\citep{CerriPOP2014b,DelSartoPRE2016} or first-order FLR corrections are included~\citep{HazeltinePOF1985,HsuPOF1986,RamosPOP2005b,CerriPOP2013}. 

\subsection{Zeroth-order solution: gyrotopic CGL pressure tensor}\label{app:subsec:CGLsolution}

At zero order, $\bPi^{(0)}$ must satisfy ${\cal L}_\Bv[\bPi^{(0)}]=0$, i.e. it will be a linear combination of the basis vector spanning the kernel of the (self-adjoint) linear operator $\mathcal{L}_B$. 
Any linear combination of the the identity, $\bI$, and of the projector along the magnetic field direction, $\bv\bv$, i.e., $\bPi^{(0)}=p_1\bI+p_2\bv\bv$, is a zeroth-order solution. 
Defining the parallel and perpendicular pressures as $\pperp=p_1$ and $\ppara=p_1+p_2$, we recover the gyrotropic CGL pressure tensor~\citep{CGL1956}:
%%%%%%%%%%%%%%%%%%%%%%%%%%%%%%
\begin{equation}\label{app:bPizero}
\bPi^{(0)}\, =\, \pperp\btau\, +\, \ppara\bv\bv\,.
\end{equation}
%%%%%%%%%%%%%%%%%%%%%%%%%%%%%%
The zeroth-order solution is insensitive to the operation $\bv\to-\bv$, as anticipated.
Note that the equation for $n=0$, and thus its solution $\bPi_\alpha^{(0)}$, does not depend on the velocity field $\uv$ or on the heat flux tensor ${\bf Q}$, so the only information that we need is the direction of the magnetic field, $\bv$.
Finally, note that there is an interesting consequence of this solution in an ordering for which $\omega/\Omegaca\ll1$: because of the gyrofrequency is inversely proportional to the species' mass, $\Omegaca\propto1/\ma$, within a low-frequency dynamics we expect the lighter species (e.g., the electrons) to be naturally found very close to a gyrotropic state\footnote{This might not be true everywhere, e.g., if processes such as reconnection are involved~\citep[see, e.g.,][]{ScudderDaughtonJGRA2008,AnuaiPOP2013}.}.

\subsection{First-order solution: FLR corrections and dynamic equations for $\ppara$ and $\pperp$}

Before proceeding in the solution of the first-order equation in the perturbative expansion, (\ref{eq:app:PIfirst-eq}), we recast it in a form that is invariant under the operation $\bv\to-\bv$. In this way, we solve it only once for a solution $\bPi^{(1)}$ that encodes both $\bPi_+^{(1)}$ and $\bPi_-^{(1)}$. At this stage, we need to take into account the fact that $\bQ^{(0)}$ changes sign when we reverse the direction of $\Bv$~\citep[see e.g.,][]{GoswamiPOP2005}. Therefore, we introduce a coefficient that takes into account the relative orientation of the magnetic field with respect to the  coordinate axes, $s_m\equiv{\rm sign}[\bv\cdot\ev_m]={\rm sign}[b_m]$ (such that $s_m^{-1} = s_m$), where $\ev_m$ is the unit vector along the $m$-axis of the reference system. 
The invariant equation now reads~\citep{CerriPOP2013}
%%%%%%%%%%%%%%%%%%%%%%%%%%%%%%%%%%%%%%%%
\[
 \Big(\epsilon_{ilm} \Pi_{lj}^{(1)}\, +\, \epsilon_{jlm} \Pi_{li}^{(1)}\Big)b_m\, =
\]
\begin{equation}\label{eq:app:pt1order-c}
 \frac{s_m\sigma_\alpha}{B}\left[\frac{{\rm d}\,\Pi_{ij}^{(0)}}{{\rm d}t}\, +\, 
 \Pi_{ij}^{(0)}\frac{\partial\,u_k}{\partial x_k}\, +\,
 \Pi_{ik}^{(0)}\frac{\partial\,u_j}{\partial x_k}\, +\, 
 \Pi_{jk}^{(0)}\frac{\partial\,u_i}{\partial x_k}\right]\, +\,
 \frac{\sigma_\alpha}{B}\frac{\partial\,Q_{ijk}^{(0)}}{\partial x_k}\,.
\end{equation}
%%%%%%%%%%%%%%%%%%%%%%%%%%%%%%%%%%%%%%%%

By evaluating every term in the above equation \citep[see, e.g.,][]{CerriPOP2013}, one eventually gets the dynamic equations for the zeroth-order pressure components,
%%%%%%%%%%%%%%%%%%%%%%%%%%%%%%%%%%%%%%%%
\begin{equation}\label{eq:p_para_App}
 \frac{\partial\,\pparaa}{\partial t}\, +\, \bnabla\cdot\big(\pparaa\uav\big)\, 
 +\, 2\,\pparaa\big(\bv\bv:\bnabla\uav\big)\, 
 +\, \bnabla\cdot\big(\qapara\bv\big)\, -\, 2\,\qaperp\big(\bnabla\cdot\bv\big)\, =\, 0\,,
\end{equation}
%%%%%%%%%%%%%%%%%%%%%%%%%%%%%%%%%%%%%%%%
\begin{equation}\label{eq:p_perp_App}
 \frac{\partial\,\pperpa}{\partial t}\, +\, \bnabla\cdot\big(\pperpa\uav\big)\, 
 +\, \pperpa\big(\btau:\bnabla\uav\big)\, 
 +\, \bnabla\cdot\big(\qaperp\bv\big)\, +\, \qaperp\big(\bnabla\cdot\bv\big)\, =\, 0\,,
\end{equation}
%%%%%%%%%%%%%%%%%%%%%%%%%%%%%%%%%%%%%%%%
and the expressions for the components of $\bPia^{(1)}$,
%%%%%%%%%%%%%%%%%%%%%%%%%%%%%%%%%%%%%%%%%%%%%%%%%%%%%
\begin{equation}\label{eq:app:P1xx}
 \Pi_{\alpha,xx}^{(1)}\, =\, -\,\Pi_{\alpha,yy}^{(1)}\, =\, 
 -\,\frac{s_3\sigma_\alpha}{2}\,\frac{\pperpa}{B}
 \left(\frac{\partial\,u_{\alpha,x}}{\partial y}\, +\,
 \frac{\partial\,u_{\alpha,y}}{\partial x}\right)
\end{equation}
%%%%%%%%%%%%%%%%%%%%%%%%%%%%%%%%%%%%%%%%%%%%%%%%%%%%%
\begin{equation}\label{eq:app:P1xy}
 \Pi_{\alpha,xy}^{(1)}\, =\, \Pi_{\alpha,yx}^{(1)}\, =\, 
 -\,\frac{s_3\sigma_\alpha}{2}\,\frac{\pperpa}{B}
 \left(\frac{\partial\,u_{\alpha,y}}{\partial y}\, -\,
 \frac{\partial\,u_{\alpha,x}}{\partial x}\right)
\end{equation}
%%%%%%%%%%%%%%%%%%%%%%%%%%%%%%%%%%%%%%%%%%%%%%%%%%%%%
\begin{equation}\label{eq:app:P1xz}
 \Pi_{\alpha,xz}^{(1)}\, =\, \Pi_{\alpha,zx}^{(1)}\, =\,
 -\,\frac{s_3\sigma_\alpha}{B}
 \left[\big(2\,\pparaa-\pperpa\big)\frac{\partial\,u_{\alpha,y}}{\partial z}\,
 +\,\pperpa\frac{\partial\,u_{\alpha,z}}{\partial y}\right]\,
 -\,\frac{\sigma_\alpha}{B}\,\frac{\partial\,\qaperp}{\partial y}
\end{equation}
%%%%%%%%%%%%%%%%%%%%%%%%%%%%%%%%%%%%%%%%%%%%%%%%%%%%%
\begin{equation}\label{eq:app:P1yz}
 \Pi_{\alpha,yz}^{(1)}\, =\, \Pi_{\alpha,zy}^{(1)}\, =\,
 \frac{s_3\sigma_\alpha}{B}\,
 \left[\big(2\,\pparaa-\pperpa\big)\frac{\partial\,u_{\alpha,x}}{\partial z}\,
 +\,\pperpa\frac{\partial\,u_{\alpha,z}}{\partial x}\right]\,
 +\,\frac{\sigma_\alpha}{B}\,\frac{\partial\,\qaperp}{\partial x}
\end{equation}
%%%%%%%%%%%%%%%%%%%%%%%%%%%%%%%%%%%%%%%%%%%%%%%%%%%%%
\begin{equation}\label{eq:app:P1zz}
 \Pi_{\alpha,zz}^{(1)}\, =\, 0\,.
\end{equation}
%%%%%%%%%%%%%%%%%%%%%%%%%%%%%%%%%%%%%%%%%%%%%%%%%%%%%
By neglecting the parallel heat fluxes, $q_\|$ and $q_\perp$, the above expressions can be compared with the classical results given in \citet{BraginskiiRPP1965} for the collisional case by setting $\eta_0=\eta_1=\eta_2=0$, $\eta_3=p_\perp/2\Omega$ and $\eta_4=p_\perp/\Omega$ in the Braginskii's gyro-viscous coefficients. 
Moreover, in our expressions there is a contribution to $\Pi_{\alpha,xz}^{(1)}$ and to $\Pi_{\alpha,yz}^{(1)}$ that is due to the pressure anisotropy, $[2(p_\|-p_\perp)/\Omega]\partial_zu_{\alpha,x}$ and $[2(p_\|-p_\perp)/\Omega]\partial_zu_{\alpha,y}$, respectively, which is missing in \citet{BraginskiiRPP1965} because of the assumed isotropic temperature, $T_\|=T_\perp=T$.
The above expressions for the FLR corrections explicitly account for the orientation of the magnetic field with respect to the $z$-axis through the $s_3$ coefficient. 

\section{Convergence of the FLR expansion to the full pressure tensor}\label{app:FLR-to-fullPI_convergence}

We expand the pressure tensor for the species $\alpha$, $\bPia$, as a power series in the small parameter $\varepsilon_\alpha\equiv\rho_\alpha/L\ll1$:
%%%%%%%%%%%%%%%%%%%%%%%%%%%%%%%%%%%%%%%%%%%%%%%%%%%%%
\begin{equation}\label{eq:Pi_expansion}
\bPia = \sum_{n=0}^\infty\varepsilon_\alpha^n\bPia^{(n)}\,,
\end{equation}
%%%%%%%%%%%%%%%%%%%%%%%%%%%%%%%%%%%%%%%%%%%%%%%%%%%%%
and we perform an equivalent expansion for the heat flux tensor, $\bQa$. Within the $e$TF ordering~\citep{CerriPOP2013}, the dimensionless $n$-th order pressure tensor equation reads
%%%%%%%%%%%%%%%%%%%%%%%%%%%%%%%%%%%%%%%%%%%%%%%%%%%%%
\begin{equation}\label{eq:n-th_equation}
 \mathcal{L}_\Bv\left[\Pi_{\alpha,ij}^{(n)}\right] = \hat{\mathcal{R}}_\uv\left[\Pi_{\alpha,ij}^{(n-1)}\right] + \mathcal{D}\left[Q_{\alpha,ij(k)}^{(n-1)}\right]\,,
\end{equation}
%%%%%%%%%%%%%%%%%%%%%%%%%%%%%%%%%%%%%%%%%%%%%%%%%%%%%
where
%%%%%%%%%%%%%%%%%%%%%%%%%%%%%%%%%%%%%%%%%%%%%%%%%%%%% 
\begin{subequations} 
\label{eq:LRD_operators}
\begin{eqnarray}
 \mathcal{L}_\Bv\left[\Pi_{\alpha,ij}^{(n)}\right] & \equiv & \left(\epsilon_{ilm}\Pi_{\alpha,lj}^{(n)}+\epsilon_{jlm}\Pi_{\alpha,li}^{(n)}\right)B_m\,,\\
 \hat{\mathcal{R}}_\uv\left[\Pi_{\alpha,ij}^{(n-1)}\right] & \equiv & s_m\sigma_\alpha\left[\frac{d\Pi_{\alpha,ij}^{(n-1)}}{dt}+\Pi_{\alpha,ij}^{(n-1)}\duakdxk + \Pi_{\alpha,ik}^{(n-1)}\duajdxk + \Pi_{\alpha,jk}^{(n-1)}\duaidxk\right]\,,\\
 \mathcal{D}\left[Q_{\alpha,ij(k)}^{(n-1)}\right] & \equiv & \sigma_\alpha\dQaijkdxk\,,
\end{eqnarray}
\end{subequations}
%%%%%%%%%%%%%%%%%%%%%%%%%%%%%%%%%%%%%%%%%%%%%%%%%%%%% 
where $\eijk$ is the Levi--Civita symbol, $\sigma_\alpha\equiv{\rm sign}(e_\alpha)$ is the sign of the electric charge of the $\alpha$ species and $s_m\equiv{\rm sign}(\bv\cdot\ev_m)$ is the relative orientation of the magnetic field with respect to the $m$-axis of the reference system ($\bv\equiv\Bv/|\Bv|$ and $\ev_m$ are the unit vectors along the magnetic field and along the $m$-axis, respectively).
We want to find an exact solution for $\bPi_{\alpha}$, i.e. a {\em convergent} series as in (\ref{eq:Pi_expansion}) that solves (\ref{eq:n-th_equation}) for all $n$.

First of all, we note that for $n=0$, the solution of (\ref{eq:n-th_equation}), which reduces to $\mathcal{L}_\Bv\left[\Pi_{\alpha,ij}^{(0)}\right]=0$, is the gyrotropic CGL pressure tensor~\citep{CGL1956}:
%%%%%%%%%%%%%%%%%%%%%%%%%%%%%%%%%%%%%%%%%%%%%%%%%%%%% 
\begin{equation}\label{eq:n0_solution}
 \bPi_\alpha^{(0)}= \pperpa\boldsymbol{\tau} + \pparaa\bv\bv
\end{equation}
%%%%%%%%%%%%%%%%%%%%%%%%%%%%%%%%%%%%%%%%%%%%%%%%%%%%%
where $\boldsymbol{\tau}\equiv{\bf I}-\bv\bv$ is the projector onto the plane perpendicular to the magnetic field. 

\subsection{Assumptions and general $n$-th order solution}

In order to find a solution of Eq.~(\ref{eq:n-th_equation}) to all orders, we first need to make four assumption on the configuration, on the energy and on the closure. 
The first is to (i) neglect the heat flux tensor.
The second is that (ii) the inhomogeneity direction, the flow direction and the magnetic field direction form a right-handed basis\footnote{Note that for incompressible flows, $\bnabla\boldsymbol{\cdot}\uv=0$, this condition correspond to the case $\Bv\boldsymbol{\times}\omegav=\Bv\boldsymbol{\times}(\bnabla\boldsymbol{\times}\uv)=0$ that has been considered by \citet{DelSartoPegoraroMNRAS2018}.}, e.g. $\uv=u_y(x)\ev_y$ and $\Bv=B_z(x)\ev_z$. 
The third assumption is (iii) stationarity, i.e. no time dependence. 
Finally, (iv) we assume that any contribution to the pressure tensor beyond the gyrotropic pressure is traceless, which means that we are considering corrections at constant thermal energy.
So, summarizing the hypothesis under which we find the solution:

\begin{itemize}
\item[(i)] ${\bf Q}^{(n)}=0$\,\, $\forall\,n$;
\item[(ii)] $\bnabla\boldsymbol{\cdot}\uv=0$ and $\Bv\boldsymbol{\times}(\bnabla\boldsymbol{\times}\uv)=0$;
\item[(iii)] $\partial/\partial t=0$;
\item[(iv)]  ${\rm Tr}[\bPi_\alpha^{(n)}]=0$\,\, $\forall\,n\geq1$.\\
\end{itemize}

Under the assumptions (i)--(iv), considering the inhomogeneity to be in $x$-direction for simplicity, the solution of (\ref{eq:n-th_equation}) $\forall n\geq1$ is:
%%%%%%%%%%%%%%%%%%%%%%%%%%%%%%%%%%%%%%%%%%%%%%%%%%%%% 
\begin{equation}\label{eq:n-th_solution}
\left\{\begin{array}{c}
 \Piaij^{(n)}=0\quad  {\rm if}\ i\neq j\,,\\
 \\
 \Piaxx^{(n)} = -\Piayy^{(n)} = \Big(\widetilde{\chi}_\alpha(x)\Big)^n\,\pperpa\,, \\
 \\
 \Piazz^{(n)} = 0\,,
\end{array}\right.
\end{equation}
%%%%%%%%%%%%%%%%%%%%%%%%%%%%%%%%%%%%%%%%%%%%%%%%%%%%% 
where we have defined the function $\widetilde{\chi}_\alpha(x)$ as
%%%%%%%%%%%%%%%%%%%%%%%%%%%%%%%%%%%%%%%%%%%%%%%%%%%%% 
\begin{equation}\label{eq:atilde}
\widetilde{\chi}_\alpha(x)\,\equiv\,-\,\sigma_\alpha\,\frac{\boldsymbol{\omega}_\alpha\cdot\bv}{2\,\Omegaca}\,=\,-\sigma_\alpha\,\frac{s_z}{2\,|\Bv|}\duaydx\,.
\end{equation}
%%%%%%%%%%%%%%%%%%%%%%%%%%%%%%%%%%%%%%%%%%%%%%%%%%%%% 
Note that, in general, $\Pizz^{(n)}$ is undetermined at each order, so we make the reasonable choice to take it nonzero only for $n=0$, i.e. $\Pizz^{(n)}=p_\|\delta_{n0}$, which then, together with the traceless condition (iv), gives us the relation $\Pi_{xx}^{(n)}+\Pi_{yy}^{(n)}=2p_\perp\delta_{n0}$.

\subsection{General $n$-th order solution: proof}

We now proceed to prove that (\ref{eq:n-th_solution}) is the solution of (\ref{eq:n-th_equation}), for all $n$. In order to do that, we are going to use the so-called mathematical induction method. Later on, we will omit the $\alpha$ index for the species for shortness.\\

\begin{itemize}

\item\underline{$n=1$}: For $n=1$, (\ref{eq:n-th_equation}) is $\mathcal{L}_\Bv\left[\Piaij^{(1)}\right] = \hat{\mathcal{R}}_\uv\left[\Piaij^{(0)}\right]$, or written in matrix form
%%%%%%%%%%%%%%%%%%%%%%%%%%%%%%%%%%%%%%%%%%%%%%%%%%%%% 
\begin{equation}\label{eq:n1_matrix_eq}
 \left(\begin{array}{ccc}
 2\Pi_{xy}^{(1)} & \Pi_{yy}^{(1)}-\Pi_{xx}^{(1)} & \Pi_{yz}^{(1)}\\
 \\
 \Pi_{yy}^{(1)}-\Pi_{xx}^{(1)} & -2\Pi_{xy}^{(1)} & -\Pi_{xz}^{(1)} \\
 \\
 \Pi_{yz}^{(1)} & -\Pi_{xz}^{(1)} & 0
 \end{array}\right) = \frac{s_3\sigma}{|\Bv|}
 \left(\begin{array}{ccc}
 \frac{d}{dt}\Pixx^{(0)} & \Pixx^{(0)}\duydx & 0\\
 \\
 \Pixx^{(0)}\duydx & -\frac{d}{dt}\Pixx^{(0)} & 0 \\
 \\
 0 & 0 & 0
 \end{array}\right)
\end{equation}
%%%%%%%%%%%%%%%%%%%%%%%%%%%%%%%%%%%%%%%%%%%%%%%%%%%%%
whose solution under our assumptions is:
%%%%%%%%%%%%%%%%%%%%%%%%%%%%%%%%%%%%%%%%%%%%%%%%%%%%% 
\begin{equation}\label{eq:n1_solution}
\left\{\begin{array}{c}
 \Piaij^{(1)}=0\quad  {\rm if}\ i\neq j\,,\\
 \\
 \Piaxx^{(1)} = -\Piayy^{(1)} = -\frac{s_z\sigma}{2|\Bv|}\duydx p_\perp \equiv\widetilde{\chi}(x)\,p_\perp\,, \\
 \\
 \Piazz^{(1)} = 0\,,
\end{array}\right.
\end{equation}
%%%%%%%%%%%%%%%%%%%%%%%%%%%%%%%%%%%%%%%%%%%%%%%%%%%%% 
where we have used the assuption (ii) and (iii) in order to have $d\Pi_{xx}^{(0)}/dt=0$: since every quantity can be function only of $x$ and the flow is along the $y$-direction due to assumption (ii), we get $\uv\cdot\nabla\Pi_{xx}^{(0)}=u_y\partial\Pi_{xx}^{(0)}/\partial y = 0$ and thus, due also to the stationariety assmption (iii), $(d/dt)\Pi_{xx}^{(0)}=(\partial/\partial t + u_y\partial/\partial y)\Pi_{xx}^{(0)}=0$.\\

\item\underline{$n=2$}: For $n=2$, (\ref{eq:n-th_equation}) is $\mathcal{L}_\Bv\left[\Piaij^{(2)}\right] = \hat{\mathcal{R}}_\uv\left[\Piaij^{(1)}\right]$, with $\Piaij^{(1)}$ given in (\ref{eq:n1_solution}). Such equation, written in matrix form reads
%%%%%%%%%%%%%%%%%%%%%%%%%%%%%%%%%%%%%%%%%%%%%%%%%%%%% 
\begin{equation}\label{eq:n2_matrix_eq}
 \left(\begin{array}{ccc}
 2\Pixy^{(2)} & \Piyy^{(2)}-\Pixx^{(2)} & \Piyz^{(2)}\\
 \\
 \Piyy^{(2)}-\Pixx^{(2)} & -2\Pixy^{(2)} & -\Pixz^{(2)} \\
 \\
 \Piyz^{(2)} & -\Pixz^{(2)} & 0
 \end{array}\right) = \frac{s_z\sigma}{|\Bv|}
 \left(\begin{array}{ccc}
 \frac{d}{dt}\Pixx^{(1)} & \Pixx^{(1)}\duydx & 0\\
 \\
 \Pixx^{(1)}\duydx & -\frac{d}{dt}\Pixx^{(1)} & 0 \\
 \\
 0 & 0 & 0
 \end{array}\right)
\end{equation}
%%%%%%%%%%%%%%%%%%%%%%%%%%%%%%%%%%%%%%%%%%%%%%%%%%%%%
whose solution, using again the fact that $d\Pi_{xx}^{(1)}/dt=0$, is:
%%%%%%%%%%%%%%%%%%%%%%%%%%%%%%%%%%%%%%%%%%%%%%%%%%%%% 
\begin{equation}\label{eq:n2_solution}
\left\{\begin{array}{c}
 \Piaij^{(2)}=0\quad  {\rm if}\ i\neq j\,,\\
 \\
 \Piaxx^{(2)} = -\Piayy^{(2)} = \frac{1}{4|\Bv|^2}\left(\duydx\right)^2 p_\perp \equiv\Big(\widetilde{\chi}(x)\Big)^2 p_\perp\,, \\
 \\
 \Piazz^{(2)} = 0\,,
\end{array}\right.
\end{equation}
%%%%%%%%%%%%%%%%%%%%%%%%%%%%%%%%%%%%%%%%%%%%%%%%%%%%% 
where we used the fact that $s_z^2=1$ and $\sigma^2=1$.\\

\item\underline{Inductive step}: We now assume that (\ref{eq:n-th_solution}) is the correct $n$-th order solution and we want to solve (\ref{eq:n-th_equation}) for the ($n+1$)-th order. That is, $\mathcal{L}_\Bv\left[\Piaij^{(n+1)}\right] = \hat{\mathcal{R}}_\uv\left[\Piaij^{(n)}\right]$, which in matrix form reads
%%%%%%%%%%%%%%%%%%%%%%%%%%%%%%%%%%%%%%%%%%%%%%%%%%%%% 
\begin{equation}\label{eq:n+1_matrix_eq}
 \left(\begin{array}{ccc}
 2\Pixy^{(n+1)} & \Piyy^{(n+1)}-\Pixx^{(n+1)} & \Piyz^{(n+1)}\\
 \\
 \Piyy^{(n+1)}-\Pixx^{(n+1)} & -2\Pixy^{(n+1)} & -\Pixz^{(n+1)} \\
 \\
 \Piyz^{(n+1)} & -\Pixz^{(n+1)} & 0
 \end{array}\right) = \frac{s_z\sigma}{|\Bv|}
 \left(\begin{array}{ccc}
 \frac{d}{dt}\Pixx^{(n)} & \Pixx^{(n)}\duydx & 0\\
 \\
 \Pixx^{(n)}\duydx & -\frac{d}{dt}\Pixx^{(n)} & 0 \\
 \\
 0 & 0 & 0
 \end{array}\right)
\end{equation}
%%%%%%%%%%%%%%%%%%%%%%%%%%%%%%%%%%%%%%%%%%%%%%%%%%%%%
whose solution, using again the fact that our assumptions are such that $d\Pi_{xx}^{(n)}/dt=0$, is:
%%%%%%%%%%%%%%%%%%%%%%%%%%%%%%%%%%%%%%%%%%%%%%%%%%%%% 
\begin{equation}\label{eq:n+1_solution}
\left\{\begin{array}{c}
 \Piaij^{(n+1)}=0\quad  {\rm if}\ i\neq j\,,\\
 \\
 \Piaxx^{(n+1)} = -\Piayy^{(n+1)} = \left(-\frac{s_z\sigma}{2|\Bv|}\duydx\right)^{n+1} p_\perp \equiv\Big(\widetilde{\chi}(x)\Big)^{n+1} p_\perp\,, \\
 \\
 \Piazz^{(n+1)} = 0\,,
\end{array}\right.
\end{equation}
%%%%%%%%%%%%%%%%%%%%%%%%%%%%%%%%%%%%%%%%%%%%%%%%%%%%%
which finally proves the thesis.\flushright$\square$

\end{itemize}

\subsection{Summabiliy, convergence and stability of the complete pressure tensor}

Now that we have proved the expression for the general $n$-th order solution of (\ref{eq:n-th_equation}), we want to go back from the FLR expansion to the full pressure tensor, (\ref{eq:Pi_expansion}). In order to be able to do that, the series must be summable and it should converge.\\
If we put all the FLR contributions together, the full pressure tensor components are:
%%%%%%%%%%%%%%%%%%%%%%%%%%%%%%%%%%%%%%%%%%%%%%%%%%%%% 
\begin{equation}\label{eq:Pi_ij}
 \Piij = 0\qquad{\rm if}\,\,i\neq j\,,
\end{equation}
%%%%%%%%%%%%%%%%%%%%%%%%%%%%%%%%%%%%%%%%%%%%%%%%%%%%%
%%%%%%%%%%%%%%%%%%%%%%%%%%%%%%%%%%%%%%%%%%%%%%%%%%%%% 
\begin{equation}\label{eq:Pi_zz0}
 \Pizz = p_\|\,,
\end{equation}
%%%%%%%%%%%%%%%%%%%%%%%%%%%%%%%%%%%%%%%%%%%%%%%%%%%%%
%%%%%%%%%%%%%%%%%%%%%%%%%%%%%%%%%%%%%%%%%%%%%%%%%%%%% 
\begin{equation}\label{eq:Pi_xx}
 \Pixx = \left[1+\widetilde{\chi}+\widetilde{\chi}^2+\dots\right]p_\perp = 
 \left[1 + \widetilde{\chi}\sum_{n=0}^\infty\left(\widetilde{\chi}\right)^n\right]p_\perp\,,
\end{equation}
%%%%%%%%%%%%%%%%%%%%%%%%%%%%%%%%%%%%%%%%%%%%%%%%%%%%%
%%%%%%%%%%%%%%%%%%%%%%%%%%%%%%%%%%%%%%%%%%%%%%%%%%%%% 
\begin{equation}\label{eq:Pi_yy}
 \Piyy = \left[1-\widetilde{\chi}-\widetilde{\chi}^2-\dots\right]p_\perp = 
 \left[1-\widetilde{\chi}\sum_{n=0}^\infty\left(\widetilde{\chi}\right)^n\right]p_\perp\,,
\end{equation}
%%%%%%%%%%%%%%%%%%%%%%%%%%%%%%%%%%%%%%%%%%%%%%%%%%%%%
so the main request for {\em absolute convergence} is that the geometric series $\sum_n|\widetilde{\chi}|^n$ converge, which is true if and only if
%%%%%%%%%%%%%%%%%%%%%%%%%%%%%%%%%%%%%%%%%%%%%%%%%%%%% 
\begin{equation}\label{eq:conv_cond}
 |\widetilde{\chi}(x)| < 1\qquad\Longleftrightarrow\qquad 
 \left|\boldsymbol{\omega}\cdot\bv\right| < 2\,\Omega_c\,\quad\forall\,x\,,
\end{equation}
%%%%%%%%%%%%%%%%%%%%%%%%%%%%%%%%%%%%%%%%%%%%%%%%%%%%%
which is the {\em absolute convergence condition} from the mathematical point of view ad represent a limit on the shear strength. If the condition (\ref{eq:conv_cond}) holds, then the resulting diagonal components of the pressure tensor are
%%%%%%%%%%%%%%%%%%%%%%%%%%%%%%%%%%%%%%%%%%%%%%%%%%%%% 
\begin{equation}\label{eq:Pi_xx_conv}
 \Pixx = \left(1+\frac{\widetilde{\chi}(x)}{1-\widetilde{\chi}(x)}\right)p_\perp\,,
\end{equation}
%%%%%%%%%%%%%%%%%%%%%%%%%%%%%%%%%%%%%%%%%%%%%%%%%%%%%
%%%%%%%%%%%%%%%%%%%%%%%%%%%%%%%%%%%%%%%%%%%%%%%%%%%%% 
\begin{equation}\label{eq:Pi_yy_conv}
 \Piyy = \left(1-\frac{\widetilde{\chi}(x)}{1-\widetilde{\chi}(x)}\right)p_\perp\,,
\end{equation}
%%%%%%%%%%%%%%%%%%%%%%%%%%%%%%%%%%%%%%%%%%%%%%%%%%%%%
%%%%%%%%%%%%%%%%%%%%%%%%%%%%%%%%%%%%%%%%%%%%%%%%%%%%% 
\begin{equation}\label{eq:Pi_zz_bis}
 \Pizz = p_\|\,.
\end{equation}
%%%%%%%%%%%%%%%%%%%%%%%%%%%%%%%%%%%%%%%%%%%%%%%%%%%%%
However, since the components of the (diagonal) pressure tensor cannot be negative in order to have a physical meaning, the function $\widetilde{\chi}(x)$ -- and thus the shear strength $du_y/dx$ -- has to fulfill the positivity condition. This request gives a {\em physical condition} on the shear strength which reads
%%%%%%%%%%%%%%%%%%%%%%%%%%%%%%%%%%%%%%%%%%%%%%%%%%%%% 
\begin{equation}\label{eq:phys_cond}
 \widetilde{\chi}(x)\,\leq\,\frac{1}{2}\qquad\Longleftrightarrow\qquad
 \boldsymbol{\omega}\cdot\bv\,\geq\,-\,\Omega_c\,,
\end{equation}
%%%%%%%%%%%%%%%%%%%%%%%%%%%%%%%%%%%%%%%%%%%%%%%%%%%%%
where now, in principle, the shear can be as negative as one wishes, without no limitations. If we put together the physical condition (\ref{eq:phys_cond}) and the mathematical condition (\ref{eq:conv_cond}), gives the {\em asymmetric} condition
%%%%%%%%%%%%%%%%%%%%%%%%%%%%%%%%%%%%%%%%%%%%%%%%%%%%% 
\begin{equation}\label{eq:asymm_cond}
 -\,1\,<\,\widetilde{\chi}(x)\,\leq\,\frac{1}{2}\qquad\Longleftrightarrow\qquad
 -\,\Omega_c\,\leq\,\boldsymbol{\omega}\cdot\bv\,<\,2\,\Omega_c\,\,\forall\,x\,,
\end{equation}
%%%%%%%%%%%%%%%%%%%%%%%%%%%%%%%%%%%%%%%%%%%%%%%%%%%%%
The condition above is also a {\em stability condition} for the shear-flow configuration. In fact, that is in agreement with \cite{DelSartoPRE2016}, where the stability condition is found to be $\Omega'\equiv\Omega+\partial_x u_y\geq0$, which translated in our notation correspond to $\widetilde{\chi}(x)\leq1/2$.\footnote{Note that the condition $\widetilde{\chi}>-1$ in (\ref{eq:asymm_cond}) originates from the fact that we are requiring that the pressure tensor $\Pi$ can be expanded in an infinite series of a small parameter $\varepsilon$, (\ref{eq:Pi_expansion}), and that the resulting contributions $\Pi^{(n)}$ should converge again to $\Pi$ when ``summed back''. However those assumptions are not made when dealing with the pressure tensor equation, so the upper bound $\boldsymbol{\omega}\cdot\bv<2\,\Omega_c$ does enter the full pressure-tensor case~\citep[see][]{CerriPOP2014b}.}

\bibliographystyle{jpp}
\bibliography{biblio}

\end{document}